\PassOptionsToPackage{table,xcdraw,dvipsnames}{xcolor}
\documentclass[acmsmall,screen,nonacm]{acmart}
\usepackage[utf8]{inputenc}
\usepackage[T1]{fontenc} 
\usepackage{xspace}
\usepackage{enumitem}
\usepackage{textcomp}
\usepackage{listings}
\usepackage{multirow}
\usepackage{flafter} 
\usepackage{subcaption}

\usepackage{tabularx}
\usepackage{tcolorbox}
\usepackage{tikz}
\usepackage[export]{adjustbox}
\usepackage{tabularray}
\UseTblrLibrary{varwidth}
\UseTblrLibrary{counter} 

\definecolor{codecolor}{RGB}{14, 23, 57}
\newcommand{\code}[1]{{\color{codecolor}\texttt{#1}}\xspace}

\newcommand{\rsq}[1]{{\itshape{#1}}}

\newcommand{\CC}{C{\texttt{++}}}

\newcounter{surveyq}[table]
\renewcommand{\thesurveyq}{\thetable\alph{surveyq}}
\newcommand{\qlabel}[1]{%
  \refstepcounter{surveyq}
  \label{#1}
  \textbf{\thesurveyq}
}

\newcounter{participantid}
\newcommand{\plink}[1]{\hyperref[figure:method:demo]{P#1}}

\newcommand{\ilquote}[2]{``\textit{#1}''~(\plink{#2})}

\newcommand{\ilquotenonum}[1]{``\textit{#1}''}

\NewDocumentEnvironment{pquote}{m}
{
    \begin{quote}\itshape
}
{
    \noindent \null \hfill {\normalfont (\plink{#1}) }
    \end{quote}
}

\newcommand{\internote}[1]{[\textit{#1}]}

\definecolor{boxcolor}{RGB}{238, 223, 204} %

\newcommand{\tzbars}[5]{
    \begin{tikzpicture}[baseline={(current bounding box.center)}]

            \pgfmathsetmacro{\H}{0.25cm}
            \pgfmathsetmacro{\W}{1.5/100}
            \pgfmathsetmacro{\CurrentX}{0}

            \pgfmathsetmacro{\SegmentLength}{#5 * \W}
            \draw [fill=l1, draw=none] (\CurrentX, 0) 
                  rectangle (\CurrentX + \SegmentLength, 0.25cm);
            \pgfmathsetmacro{\CurrentX}{\CurrentX + \SegmentLength}

            \pgfmathsetmacro{\SegmentLength}{#4 * \W}
            \draw [fill=l2, draw=none] (\CurrentX, 0) 
                  rectangle (\CurrentX + \SegmentLength, 0.25cm);
            \pgfmathsetmacro{\CurrentX}{\CurrentX + \SegmentLength}

            \pgfmathsetmacro{\SegmentLength}{#3 * \W}
            \draw [fill=mid, draw=none] (\CurrentX, 0) 
                  rectangle (\CurrentX + \SegmentLength, 0.25cm);
            \pgfmathsetmacro{\CurrentX}{\CurrentX + \SegmentLength}
                    
            \pgfmathsetmacro{\SegmentLength}{#2 * \W}
            \draw [fill=r1, draw=none] (\CurrentX, 0) 
                  rectangle (\CurrentX + \SegmentLength, 0.25cm);
            \pgfmathsetmacro{\CurrentX}{\CurrentX + \SegmentLength}

            \pgfmathsetmacro{\SegmentLength}{#1 * \W}
            \draw [fill=r2, draw=none] (\CurrentX, 0) 
                  rectangle (\CurrentX + \SegmentLength, 0.25cm);
            \pgfmathsetmacro{\CurrentX}{\CurrentX + \SegmentLength}

    \end{tikzpicture}
}

\newcommand{\tzbarsunsure}[6]{
    \begin{tikzpicture}[baseline={(current bounding box.center)}]
        \pgfmathsetmacro{\H}{0.25cm}
        \pgfmathsetmacro{\W}{1.5/100}
        \pgfmathsetmacro{\CurrentX}{0}
        \pgfmathsetmacro{\SegmentLength}{#6 * \W}
        \draw [fill=l1, draw=none] (\CurrentX, 0) 
              rectangle (\CurrentX + \SegmentLength, 0.25cm);
        \pgfmathsetmacro{\CurrentX}{\CurrentX + \SegmentLength}

        \pgfmathsetmacro{\SegmentLength}{#5 * \W}
        \draw [fill=l2.1, draw=none] (\CurrentX, 0) 
              rectangle (\CurrentX + \SegmentLength, 0.25cm);
        \pgfmathsetmacro{\CurrentX}{\CurrentX + \SegmentLength}

        \pgfmathsetmacro{\SegmentLength}{#4 * \W}
        \draw [fill=l2.2, draw=none] (\CurrentX, 0) 
              rectangle (\CurrentX + \SegmentLength, 0.25cm);
        \pgfmathsetmacro{\CurrentX}{\CurrentX + \SegmentLength}

        \pgfmathsetmacro{\SegmentLength}{#3 * \W}
        \draw [fill=mid, draw=none] (\CurrentX, 0) 
              rectangle (\CurrentX + \SegmentLength, 0.25cm);
        \pgfmathsetmacro{\CurrentX}{\CurrentX + \SegmentLength}
                
        \pgfmathsetmacro{\SegmentLength}{#2 * \W}
        \draw [fill=r1, draw=none] (\CurrentX, 0) 
              rectangle (\CurrentX + \SegmentLength, 0.25cm);
        \pgfmathsetmacro{\CurrentX}{\CurrentX + \SegmentLength}

        \pgfmathsetmacro{\SegmentLength}{#1 * \W}
        \draw [fill=r2, draw=none] (\CurrentX, 0) 
              rectangle (\CurrentX + \SegmentLength, 0.25cm);
        \pgfmathsetmacro{\CurrentX}{\CurrentX + \SegmentLength}
    \end{tikzpicture}
}

\newcommand{\tzplotbars}[6]{%
    \begin{adjustbox}{valign=c}
        \begin{tikzpicture}[
            baseline={(current bounding box.center)},
            block/.style={
            minimum width={width("(100\%)")}}
        ]
        \node[anchor=west, inner sep=0pt] (bar) at (0, 0) {%
            \tzbars{#1}{#2}{#3}{#4}{#5}%
        };
        \node[block, anchor=east] at (bar.west) [xshift=0pt, yshift=0pt] {\strut\fpeval{round(#4 + #5)}\%};
        \node[block, anchor=west] at (bar.east) [xshift=0pt, yshift=0pt] {\strut\fpeval{round(#1 + #2)}\%}; 
        \node[block, anchor=west] at ([xshift=0.7cm, yshift=0pt]bar.east) {\strut(#6)}; 
        \end{tikzpicture}%
    \end{adjustbox}
}

\newcommand{\tzplotbarsunsure}[7]{%
    \begin{adjustbox}{valign=c}
        \begin{tikzpicture}[
            baseline={(current bounding box.center)},
            block/.style={
            minimum width={width("(100\%)")}}
        ]
            \node[anchor=west, inner sep=0pt] (bar) at (0, 0) {%
                \tzbarsunsure{#1}{#2}{#3}{#4}{#5}{#6}%
            };
        \node[block, anchor=east] at (bar.west) [xshift=0pt, yshift=0pt] {\strut\fpeval{round(#5 + #6)}\%};
        \node[block, anchor=west] at (bar.east) [xshift=0pt, yshift=0pt] {\strut\fpeval{round(#1 + #2)}\%}; 
        \node[block, anchor=west] at ([xshift=1cm, yshift=0pt]bar.east) {\strut(#7)}; 
        \end{tikzpicture}%
    \end{adjustbox}
}

\DeclareRobustCommand{\tlink}[2]{\hyperref[#2]{#1~\getrefnumber{#2}}}
\newcommand{\circlesize}{0.8ex}

\newcommand*\checkedbox[1][\circlesize]{%
\begin{tikzpicture}
    \draw[thick] (0,0) rectangle (2,2);
    
    \draw[green, thick] (0.5, 1) -- (1, 0.5) -- (1.5, 1.5);
\end{tikzpicture}
}
\newcommand*\unchecked[1][\circlesize]{%
\begin{tikzpicture}
    \draw[thick] (0,0) rectangle (2,2);
    
    \draw[green, thick] (0.5, 1) -- (1, 0.5) -- (1.5, 1.5);
\end{tikzpicture}
}

\definecolor{l1}{RGB}{43,123,182}
\definecolor{l2}{RGB}{172,217,233}
\definecolor{mid}{RGB}{255,255,190}
\definecolor{r1}{RGB}{253,174,96}
\definecolor{r2}{RGB}{214,23,27}
\newcommand{\cbox}[1]{\tikz[baseline=1pt]{\path[draw=#1,fill=#1] (0,0) rectangle (.25cm,.25cm);}}
\definecolor{l2.1}{RGB}{146,190,219}
\definecolor{l2.2}{RGB}{222,244,249}
\begin{document}
\title{A Mixed Methods Study on the Implications of Unsafe Rust for Interoperation, Encapsulation, and Tooling}
\author{Ian McCormack}
\email{icmccorm@cs.cmu.edu}
\affiliation{%
 \institution{Carnegie Mellon University}
 \city{Pittsburgh}
 \state{Pennsylvania}
 \country{USA}
}
\author{Tomás Dougan}
\email{tomas_dougan@alumni.brown.edu}
\affiliation{%
 \institution{Brown University}
 \city{Pittsburgh}
 \state{Pennsylvania}
 \country{USA}
}
\author{Sam Estep}
\email{estep@cmu.edu}
\affiliation{%
 \institution{Carnegie Mellon University}
 \city{Pittsburgh}
 \state{Pennsylvania}
 \country{USA}
}
\author{Hanan Hibshi}
\email{hhibshi@cmu.edu}
\affiliation{%
 \institution{Carnegie Mellon University}
 \city{Pittsburgh}
 \state{Pennsylvania}
 \country{USA}
}
\affiliation{%
 \institution{King Abdulaziz University}
 \city{Jeddah}
 \country{Saudi Arabia}
}

\author{Jonathan Aldrich}
\email{jonathan.aldrich@cs.cmu.edu}
\affiliation{%
 \institution{Carnegie Mellon University}
 \city{Pittsburgh}
 \state{Pennsylvania}
 \country{USA}
}
\author{Joshua Sunshine}
\email{sunshine@cs.cmu.edu}
\affiliation{%
 \institution{Carnegie Mellon University}
 \city{Pittsburgh}
 \state{Pennsylvania}
 \country{USA}
}

\renewcommand{\footnotemark}{\mbox{}} 
\titlenote{
  This material is based upon work supported by the U.S. Department of Defense  
  under Grant No. H98230-23-C-0274 and the 
  \grantsponsor{GS100000001}{National Science
    Foundation}{http://dx.doi.org/10.13039/100000001} under Grant
  Nos.~\grantnum{GS100000001}{CCF-1901033}, \grantnum{GS100000001}{CCF-2447499}, \grantnum{GS100000001}{CCF-2339830}, \grantnum{GS100000001}{DGE1745016}, and \grantnum{GS100000001}{DGE2140739}.
  Any opinions, findings, conclusions, or recommendations expressed in this material are those of the authors and do not necessarily reflect the views of the Department of Defense or the National Science Foundation.
}

\begin{abstract}
The Rust programming language restricts aliasing to provide static safety guarantees. However, in certain situations, developers need to bypass these guarantees by using a set of ``unsafe'' features. When these features are used incorrectly, they can reintroduce the types of safety issues that Rust was designed to prevent. We seek to understand how current development tools can be improved to better assist developers who find it necessary to interact with unsafe code. To that end, we study how developers reason about foreign function calls, the limitations of the tools that they currently use, their motivations for using unsafe code, and how they reason about encapsulating it. We conducted a mixed methods investigation consisting of semi-structured interviews with 19 developers, followed by a survey that reached an additional 160 developers. Our participants were motivated to use unsafe code when they perceived that there was no alternative, and most claimed that they would avoid using it. However, limited tooling support for foreign function calls made participants uncertain about their design choices, and certain foreign aliasing and concurrency patterns were difficult to encapsulate. To overcome these challenges, developers will need analysis tools that can find Rust-specific forms of undefined behavior within multilanguage applications.
\end{abstract}
\maketitle

\section{Introduction}
\label{section:intro}
The Rust programming language has made rapid headway as a safe alternative for systems programming due to its static safety guarantees. However, it is still the case that critical systems are written in languages that do not provide inherent safety guarantees~\cite{rust_kernel_policy, google_jansens23}. Rust developers who interoperate with these applications must use a special subset of ``unsafe'' features, including calling foreign functions and accessing memory through raw ``C-style'' pointers. These features can only be used within a block or function marked with the \code{unsafe} keyword. 

To minimize the risks of unsafe code, the Rust community advocates for keeping it minimal, well documented, and hidden beneath a safe interface~\cite{astrauskas20}. However, these practices alone are not enough to ensure that Rust delivers on its promise of static safety. Errors in safe encapsulations of unsafe implementations have been a significant source of bugs and security vulnerabilities in the Rust ecosystem~\cite{qin20,xu21,zheng23}. Applications that rely on interoperation or direct access to system resources may expose developers to sources of unsafety that cannot easily be minimized. Making matters worse, Rust's static and dynamic semantics are still evolving~\cite{treeborrows, polonius_update}, so few tools are capable of fully testing or verifying these applications. 

Interoperation is a common use case for unsafe code~\cite{evans20, astrauskas20, fulton21, holtervennhoff23}, but most Rust-specific development tools lack support for foreign function calls. Safe interoperation will be crucial for Rust's future, since both industry and regulatory stakeholders have advocated for adopting Rust as a replacement for existing systems languages~\cite{tractor, rebert_kern24}. Automated translation tools are increasingly effective~\cite{larsen18}, but for the moment, interoperation is the most practical method for adopting Rust within existing, large-scale systems~\cite{rebert_kern24}. It can be challenging to use Rust correctly in this context, because it has significant differences from other languages~\cite{holtervennhoff23, mirilli}. A detailed account of these challenges is missing, and we need this to be able to develop effective tooling for safe interoperation. To provide this perspective, we ask the following research questions:

\vskip 0.5\baselineskip

\noindent\textbf{RQ1 (Interoperation).}
\\
\rsq{How do Rust developers reason about memory safety across foreign function boundaries?}
\vskip 0.5\baselineskip
\noindent\textbf{RQ2 (Tooling).} 
\\
\rsq{What tools do Rust developers use when contributing to applications that include unsafe code, and how could tooling be improved?}
\vskip 0.5\baselineskip

Developers who interoperate with Rust will need to use a variety of unsafe features in addition to foreign function calls to be able to reconcile the differences between Rust and other languages. Having a broader understanding of what motivates Rust developers to use unsafe code in any capacity and how they reason about soundness will be helpful to identify if there are factors beyond these differences that impact the safety of multilanguage applications. However, most large-scale studies on these topics have analyzed patterns in source code instead of directly engaging with developers~\cite{evans20, astrauskas20}. Interviews and surveys have provided a missing qualitative perspective, but tend to have a limited focus on unsafe code~\cite{fulton21}, a small sample size~\cite{evans20}, or different goals~\cite{holtervennhoff23}. In addition, to serve as guidance for the design of future tools, we seek to generalize prior qualitative results to a broader population of developers. To that end, we ask the following additional research questions:

\vskip 0.5\baselineskip
\noindent\textbf{RQ3 (Motivations).}
\\
\rsq{What are Rust developers' motivations for using unsafe code?}
\vskip 0.5\baselineskip
\noindent\textbf{RQ4 (Encapsulation).}
\\
\rsq{How do Rust developers reason about encapsulating unsafe code?}
\vskip 0.5\baselineskip

\paragraph{Contribution} We provide the results of an exploratory, mixed method study that addresses the research questions above. We conducted semi-structured interviews with 19 developers and used the results to create a community survey that had 160 valid responses. Most of our participants used unsafe code to call foreign functions, and they described several aliasing and concurrency patterns that made interoperation difficult. Many developers had used Miri~\cite{miri}, a Rust interpreter, to find bugs, but most were deterred from using it again due to its slow performance and lack of support for key features. Participants prioritized keeping unsafe code minimal, encapsulated, and well documented, but rarely audited their dependencies and relied on ad hoc reasoning to justify their decisions. Expanded tooling support for multilanguage applications and improved documentation will be necessary to support developers in maintaining Rust's safety guarantees across foreign function boundaries. 

\paragraph{Outline} In \tlink{Section}{section:background}, we describe the role of unsafe code in the Rust ecosystem and the challenges associated with interoperating with other languages. We describe our methodology in \tlink{Section}{section:methods}, and we indicate threats to validity in \tlink{Section}{section:threats}. We present our results in \tlink{Section}{section:results} and compare our findings with prior work in \tlink{Section}{section:related}. We discuss the implications of our results in \tlink{Section}{section:discussion} before concluding in \tlink{Section}{section:conclusion}. Our replication package is available via Zenodo~\cite{dataset}.

\section{Background}
\label{section:background}
Rust's safety guarantees are based on the notion of ownership. Values with \textit{copy} semantics do not have an owner and can be copied freely, while values with \textit{move} semantics have a unique owner. Assignment transfers ownership from one alias to another, but ownership can also be \textit{borrowed} by creating a reference. Each reference has a \textit{lifetime}, which is the duration of the program where it is used. Rust's \textit{borrow checker} enforces that a value can have either a unique mutable reference or many immutable references, but not both simultaneously~\cite{crichton20}. A value's type can implement \textit{traits} that define how it behaves. Certain traits affect a value's capability to be owned and borrowed. Values with the \code{Copy} trait have copy semantics, values that are \code{Send} can be moved between threads, and values that are \code{Sync} can be borrowed by multiple threads.

Rust's aliasing rules are conservative and are not universally compatible with certain idioms of systems programming. When these restrictions become burdensome, developers can use the \code{unsafe} keyword to enable features that are not constrained by the borrow checker. Within an unsafe context, users can dereference raw pointers, execute inline assembly instructions, modify static mutable state, implement unsafe traits, and call unsafe functions~\cite{rustbook}. There are a variety of uses for these features~\cite{astrauskas20}. In certain situations, unsafe code can be necessary to implement design patterns that are fundamentally incompatible with the restrictions of safe references, such as shared mutable state. Foreign function calls, inline assembly, and intrinsics allow developers to leverage hardware acceleration and interact with external resources. These features may also improve performance by skipping certain run-time assertions, such as bounds checks.

However, if unsafe code is used inappropriately, it can reintroduce the same types of safety issues that Rust was designed to prevent, such as data races and accesses out-of-bounds. Rust's aliasing restrictions also introduce new forms of \textit{undefined behavior}. Safe references are constrained by the borrow checker, but they can also be cast as raw  pointers, which can only be used in unsafe contexts. The Rust compiler is allowed to assume that programs which use these features are still following Rust's aliasing rules. Programs that break these rules may be optimized incorrectly, introducing differences in behavior that can constitute security vulnerabilities.

Several best practices can help mitigate the risks of using unsafe code. When implementing an unsafe feature, developers are encouraged to follow the ``interior unsafety'' pattern by keeping it minimal, well documented, and hidden beneath a safe interface~\cite{qin20}. Ideally, this interface should make it impossible for users to trigger undefined behavior from safe code. However, certain safety properties are difficult or impossible to encode into Rust's type system without leveraging external verification tools. Miri~\cite{miri}, a Rust interpreter, is currently the only tool that can detect when unsafe operations lead to violations of Rust's newest aliasing rules, so it is typically seen as the ``de facto''~\cite{gaher24} bug-finding tool for applications that use unsafe code. Currently, Miri can detect violations of two different models of Rust's semantics. The Stacked Borrows~\cite{stackedborrows} model was the first iteration of these rules; it is enabled by default. Developers can also switch to using the newer Tree Borrows~\cite{treeborrows} model by providing a configuration flag. Neither of the two models is canonical.

These practices are standard for the Rust community, but Rust is also increasingly being adopted in interoperation with preexisting large-scale C and C++ applications, such as Chromium~\cite{google_jansens23} and the Linux Kernel~\cite{rust_kernel_policy}. These applications will expose Rust components that may be ``safe'' in isolation to significant external sources of unsafety, which will not be as easy to minimize or document. In addition to these challenges, Miri does not fully support finding undefined behavior in this context. Miri can execute and trace certain foreign functions from shared libraries, but it cannot detect aliasing violations that are triggered in foreign code, and its slow performance prevents it from being used at scale.

These obstacles will make it difficult for developers to avoid introducing undefined behavior in multilanguage Rust applications. Projects such as the DARPA's TRACTOR program~\cite{tractor} and the Rust/C++ Interoperability Initiative~\cite{rust_interop_initiative} seek to provide new development tools to assist with these challenges. We need a broader understanding of how Rust developers use unsafe code to be able to design interventions that are effective in practice, since developers who use Rust in this context will necessarily engage with a variety of unsafe operations in order to reconcile differences in semantics across language boundaries.

\section{Methodology}
\label{section:methods}
We investigated how Rust developers engage with unsafe code by adapting a sequential, exploratory mixed methods design~\cite{mixed_methods_designs}, which consisted of three phases. We describe the first phase in \tlink{Section}{sections:method:interviews}, where we used a screening survey to recruit developers for semi-structured interviews. In the second phase, after interviewing each eligible candidate, the first three authors conducted a thematic analysis~\cite{miles20} of the interview transcripts. We describe this process in \tlink{Section}{sections:method:qualitative}. We describe the third phase in \tlink{Section}{sections:method:survey}, where we used the results of our qualitative analysis to create a community survey. We chose this combination of methods because it allowed us to evaluate the generalizability of prior qualitative findings and to provide a missing qualitative perspective on interoperation and tool use. 

\begin{figure}
\caption{A sample of materials relevant to each stage of our methodology. This includes the screening survey, interview protocol, an example of a theme from our codebook, and a survey question connected to that theme. Notes \internote{in brackets} are documentation and were not presented to participants.}
\label{figure:method}
\begin{subfigure}[t]{0.48\textwidth}
\caption{The screening survey.}
\label{figure:method:screening}
\scriptsize
{
\setlist{itemsep=0.5em}
\begin{enumerate}[leftmargin=*, label=\arabic*., ref=\arabic*]
\item \label{screening:ffi-rust} Do you call foreign functions from Rust?
\item \label{screening:ffi-other} Do you call Rust functions from other languages?
\item Do you use Rust's intrinsics?
\item Do you use system calls in Rust?
\item Which of the following reference and memory container types have you converted to raw pointers or used to contain raw pointers?
\item Which of the following bug finding tools do you use with codebases containing unsafe Rust?
\item \label{screening:ffi-start} \internote{If yes to \#1} - Foreign functions that you call from Rust are written in:
\item \internote{If yes to \#1} - Your code that calls foreign Rust functions is written in:
\item \label{screening:ffi-end} \internote{If yes to \#1 or \#2} - Select the tools and methods you use to create foreign bindings to Rust, or from Rust to other languages.
\end{enumerate}
}
\end{subfigure}\hfill\begin{subfigure}[t]{0.48\textwidth}
\caption{The interview protocol.}
\label{figure:method:protocol}
\scriptsize
{
\setlist{itemsep=0.5em}
\begin{enumerate}[leftmargin=*, label=\arabic*., ref=\arabic*]
\item What have been your motivations for learning Rust and choosing to use it?
\item What do you use \code{unsafe} Rust for?
\item You mentioned using \internote{screening survey \#5} in \code{unsafe} contexts. What do you use them for? Describe how you reason about memory safety in these situations.
\item Describe a bug you faced that involved \code{unsafe} Rust.
\item You mentioned using \internote{screening survey \#9} to create foreign bindings. Describe your experiences with these methods. 
\item \label{protocol:ffi} \internote{If yes to screening survey \#1 or \#2} - How do you navigate the differences between Rust's memory model and other memory models? 

\item You mentioned using \internote{screening survey \#6}. Describe your experience using each one. 
\item Do your development tools handle all of the problems that you face writing \code{unsafe} Rust?
\end{enumerate}
}
\end{subfigure}
\begin{subfigure}{\textwidth}
\vskip 0pt
\caption{Codes for the theme "Why \code{unsafe}?", which describes participants' motivations for using unsafe code.}
\label{figure:method:codebook}
\scriptsize
\bgroup
\def\arraystretch{1.5}
\begin{tabularx}{\textwidth}{p{2.75cm}XcX}
\textbf{Code}           & \textbf{Definition}                                          & \textbf{Participant} & \textbf{Quote}                                              \\ \hline
\rowcolor[HTML]{EFEFEF}
Increase Performance & The user is choosing to use unsafe because they perceive that it would be more performant than safe code. It does not matter if they actually measured performance or not.               & P3          & \textit{"I can use unsafe to speed up my code a bit and then wrap that in some sort of safer, larger abstraction."}         \\[4.2ex]
 
Easier or More Ergonomic         & There is a stated or implied choice between using a safe pattern and using an unsafe pattern, but the participant chose the unsafe pattern due to the perception that it was easier to implement or validate.                                  & P4         & \textit{"It was simpler to use the raw allocation and deallocation API directly instead of trying to use the existing higher level primitives."}           \\
\rowcolor[HTML]{EFEFEF} No Other Choice   & When there is no safe option or existing encapsulation to accomplish something. & P10         & \textit{"I use unsafe Rust for mostly to create new data structures or types that are not really possible to express with safe Rust."}\\       
\hline
\end{tabularx}
\egroup
\end{subfigure}
\begin{subfigure}{0.9\textwidth}
\caption{A survey question with a direct correspondence to the codes shown in \tlink{Figure}{figure:method:codebook}.}
\label{figure:method:questions}
\footnotesize
Think about the situations where you have typically used \code{unsafe}. Which of the following reasons have motivated you to do so?
{
\setlist{itemsep=0.5em}
\begin{itemize}[leftmargin=*]
\item I could use a safe pattern but \code{unsafe} is faster or more space-efficient.
\item I am not aware of a safe alternative at any level of ease-of-use or performance.
\item I could use a safe pattern but \code{unsafe} is easier to implement or more ergonomic.
\item Other \internote{short answer}
\end{itemize}
}
\end{subfigure}
\end{figure}

\subsection{Interviews} 
\label{sections:method:interviews}
On May 3rd, 2023, we posted a call to participate in semi-structured interviews on the Rust subreddit (\href{https://www.reddit.com/r/rust/}{/r/rust}), the Rust Programming Language forums, and the \texttt{\#dark-arts} channel of the Rust Community Discord server. We also contacted acquaintances with relevant experience. To be eligible to participate, candidates needed to have at least one year of experience using Rust and had to have ``regularly written or edited Rust code within an unsafe block or function.'' Eligible candidates completed a series of multiple choice and short answer questions about their use of unsafe features and development tools. We provide a sample of these questions in \tlink{Figure}{figure:method:screening}. We also collected participants' years of experience using Rust and we asked them to describe their affiliation. The first and third authors collaboratively coded each affiliation as one or more of the following categories: ``Industry,'' ``Rust Team,'' ``Academia,'' and ``Open Source.''

We used participants' responses to the screening survey to guide each interview. We provide a sample of our interview protocol in \tlink{Figure}{figure:method:protocol}. Certain questions were gated based on prior responses. If an candidate answered "No" to questions \hyperref[screening:ffi-rust]{\#\getrefnumber{screening:ffi-rust}} and \hyperref[screening:ffi-other]{\#\getrefnumber{screening:ffi-other}} of \tlink{Figure}{figure:method:screening}, then they did not use Rust in multilanguage applications, so we did not ask them questions \hyperref[screening:ffi-start]{\#\getrefnumber{screening:ffi-start}-\getrefnumber{screening:ffi-end}} of the screening survey or question \hyperref[protocol:ffi]{\#\getrefnumber{protocol:ffi}} of the interview protocol, since each of these questions involve foreign functions. 

The first author conducted all 19 interviews remotely over Zoom between May and July of 2023. Participants were not required to enable their camera. We used snowball sampling~\cite{patton1990qualitative} to expand our pool of participants beyond the group that responded to our initial calls to participate. If a participant shared new information that we had not learned yet, then we prompted them to invite any acquaintances with relevant experience to complete our screening survey. We stopped recruiting new participants after we had reached saturation. Our concept of saturation corresponds to Saunders’ et al~\cite{saunders18}’s definition of \textit{data saturation} as a subjective judgement made during data collection before analysis. We knew that we had reached this point after we began to ``hear the same comments again and again''~\cite{hawking88} and had informally identified several themes from our interviews. Two themes were particularly notable at this stage; certain participants felt that Rust’s \code{Box} could be cumbersome due to its \code{noalias} semantics (\tlink{Section}{results:rq1}), and that Miri was difficult to use in multilanguage applications (\tlink{Section}{results:rq2}). Our sample size (19) is comparable to other interview studies with Rust developers~\cite{fulton21, holtervennhoff23}. We transcribed the audio recordings of the interviews using Whisper~\cite{whisper}, and the first author reviewed every transcript to correct errors and label whether the utterances came from the interviewer or the interviewee.

\subsection{Qualitative Analysis}
\label{sections:method:qualitative}
We conducted an inductive, thematic analysis of our interview transcripts following best-practices by Miles, Huberman, and Saldaña~\cite{miles20}. To begin, the first three authors conducted open coding of random samples of quotes taken from each interview. To create a sample, the first author picked random character indices within each of the interview transcripts and chose a snippet of one of the interviewees' responses near each index. If all responses at that index had been sampled before, then the first author generated a new random index and repeated this process until a new quote was found. We coded three samples of quotes and met after each sample to resolve disagreements and record new codes within a shared codebook.

Next, the first and second authors used a card-sorting~\cite{braun06} exercise to identify higher-level themes among the codes that we had generated in the first phase. We created cards for each of the quotes randomly sampled in the first phase, as well as for a set of quotes that were highlighted by the first author when reviewing each transcript. We used the themes that we created from this process to organize our codebook, and we removed redundant or irrelevant codes. The first and second authors used this codebook to conduct closed coding of seven additional random samples of quotes, meeting after coding each sample to resolve disagreements and update the codebook. We used the same process as before to create samples, but instead of treating each interview transcript separately, we sampled quotes from a single document containing the concatenated text of every interview transcript. \tlink{Table}{figure:method:codebook} shows one of the themes that we created during this process, as well as its constituent codes. 

After this refinement stage, the first and second authors divided the set of transcripts between each other to code in full, independently, using ATLAS.ti Web~\cite{atlasti}. The first author coded 11 transcripts, while the second author coded the remaining 8 transcripts. Throughout this process, both authors continued to update a shared codebook with additional codes and themes to describe new observations. After all transcripts had been coded by one of the two authors, the first author conducted an additional coding pass for every transcript and resolved each of their disagreements with the second author. The complete codebook and all coding decisions are included in our replication package.

\subsection{Survey}
\label{sections:method:survey}
We constructed the community survey after we had completed coding every interview, but before the first author had finished auditing coding decisions. Each of the first three authors independently assembled a list of survey questions, which the first author assembled into a single survey in Qualtrics~\cite{qualtrics}. Excluding consent and eligibility, the survey had 85 yes/no, multiple-choice, and Likert scale questions, which were grouped into 26 sections. To avoid priming participants to select certain responses, we randomly shuffled the order of multiple choice and yes/no options and randomly reversed the order of Likert scale items. Similar to our screening survey, certain questions were gated based on prior responses. We provide a complete copy of our survey, including its branching logic, in our Appendix~\cite{dataset}.

Questions were either descriptive or exploratory. \tlink{Figure}{figure:method:questions} shows an example of a descriptive question. Each response option corresponds to the codes shown in \tlink{Figure}{figure:method:codebook} for the theme ``Why unsafe?''. Exploratory questions arose after we had finished conducting interviews, and they did not have a direct correspondence to the themes that we identified during qualitative analysis. For example, Höltervennhoff et al.~\cite{holtervennhoff23}'s study was published concurrent to our investigation, and three of their participants mentioned using either \code{cargo-audit} or \code{cargo-deny} to detect vulnerabilities in their dependencies. None of our interview participants had mentioned using these tools, but we included a survey question asking if respondents had used them before. We also relaxed our eligibility criteria for the survey to include individuals who would have been eligible for Höltervenhoff et al.~\cite{holtervennhoff23}'s study. Candidates were eligible to participate in their study if they had committed unsafe code to GitHub, but there was no indication if they had done so regularly. Our interview participants needed to have ``regularly'' engaged with unsafe code, but our survey respondents were eligible if they had ``written, edited, read, audited, or engaged with \code{unsafe} Rust code in any way.'' 

Two of our interview participants piloted an initial draft of the survey, and we incorporated their feedback into the final version. We distributed our community survey on September 22nd, 2023 to the same acquaintances and communities where we distributed our screening survey. We also published a link to our survey in the September 28th edition of the ``This Week in Rust'' newsletter and we advertised the survey during a presentation and poster sessions at an industry partners conference hosted by our institution. At the end of the survey, participants could provide a link to their profile on either GitHub or the Rust Programming Language Forum to have a chance to receive one of two \$250 gift cards. We randomly selected two participants who had provided a link to their profile with account activity prior to the start date of the survey.

We configured Qualtrics~\cite{qualtrics} to reject multiple responses from the same individual. We also enabled reCaptcha~\cite{liu18} and RelevantID~\cite{relevantid} to detect possibly fraudulent submissions. RelevantID uses several forms of metadata, including location and IP information; none of this information was retained or visible to the investigators. In our analysis, we excluded responses that had a reCaptcha score of less than 0.5, indicating that the response was likely from a bot. For RelevantID, we excluded responses that had a fraud score greater than 0.3 or were marked as likely to be duplicate. 

\section{Ethics \& Threats to Validity} 
\label{section:threats}
\paragraph{Ethics} All procedures in this study were approved by our institution's IRB. Before each intervention, participants were read or presented a script outlining the procedures of the study and they were asked to affirm consent. Participants were reminded to avoid sharing personally identifying information in their responses, and interview participants were asked to remain in a private space. The first author reviewed the contents of the replication package and redacted all direct (e.g. name, affiliation) and indirect (e.g. position, project) personal identifiers.

\paragraph*{External Validity} Recruitment via snowball sampling and through online communities may bias our results toward certain areas of the Rust community. We attempted to maintain a diverse survey and interview population by sampling from multiple forums. However, both Reddit and ``This Week in Rust'' were overrepresented compared to the other communities where we distributed our survey. Our sample does not include developers who are not active in these communities.  We cannot determine if our interview population has a similar bias, since we did not record where these participants had first learned about our study.

The "silent" developers excluded from these populations may have different experiences and perspectives that are not represented in the current findings. Rust’s unsafe “superpowers” are typically advertised as an advanced feature that developers “might run into every once in a while, but may not use every day”~\cite{rustref}. Developers who have engaged with unsafe code may more likely to have had more experience using Rust in any capacity than those who have not used unsafe code. We suspect that committed Rust experts are overrepresented in our interview population, since many were affiliated with prominent open source projects or had experience using the language within security-critical applications. 

It is more difficult to evaluate the capabilities of our survey respondents. There was significant variance in the reported number of years of experience. However, on average, our survey population had nearly the same number of years of experience using Rust as our interview population, and more than a decade of experience within the field of software engineering. Most respondents reported behaviors that indicate non-trivial experience with Rust, such as using bug-finding or auditing tools and committing unsafe code to a published library. For this reason, we expect that our current results reflect the ``best-case scenario'' for developers' experiences using unsafe Rust. Any aspects of Rust development that our population found to be difficult seem likely to be even more challenging for the broader Rust community.

Additionally, the gift cards that we offered for compensation are most easily redeemable in North America, which may have dissuaded certain individuals from participating in either stage of our study. We did not collect location data from participants, so we cannot determine if our results are biased toward individuals from any particular area. 

\paragraph*{Construct Validity} Respondents were eligible to complete our survey if they had one year of experience and had ``engaged'' with unsafe code in any way. A subset may have been eligible without having significant practical experience with unsafe code. We mitigated this by gating questions based on whether participants had used certain features. When relevant, we compare results from the entire population with results from the subset who ``regularly'' wrote or edited unsafe code, which matches the eligibility criteria for our interviews. This criteria may have introduced self-selection bias among our interview participants, since individuals who frequently use unsafe code may be more likely to perceive that it is necessary for their purposes. We did not enable reCaptcha~\cite{liu18}, RelevantID~\cite{relevantid}, or any of Qualtrics' other safeguards when recruiting participants for interviews. These safeguards are subject to false positives~\cite{zhang22}, and may have flagged legitimate participants.

The validity of our results depends on participants' ability to accurately self-report their behavior, and what developers perceive about their use of unsafe code may differ from how it is used in practice. We did not evaluate whether participants were able to reason correctly about unsafe code; only how they perceived their usage of these features. Höltervennhoff et al.~\cite{holtervennhoff23} found that developers who use unsafe code have varying degrees of awareness about its implications for correctness; 10 of their interview participants indicated that unsafe contexts entirely disable Rust’s static safety restrictions, which is incorrect. Our qualitative results are also unavoidably influenced by the individual biases of each investigator. We tried to mitigate this by auditing coding decisions and meeting frequently to resolve disagreements. Although most of our interview protocol was written to be neutral toward unsafe code, certain questions may have biased participants toward providing responses that favored safe coding practices, such as asking participants to describe how they ``reason about memory safety'' and ``navigate the differences'' between Rust and other languages. 

\section{Results}
\label{section:results}
We report the demographics of interview and survey participants in \tlink{Section}{results:demo}. Then, we describe how these developers reason about foreign functions in \tlink{Section}{results:rq1} (RQ1), and we discuss the tooling they use in \tlink{Section}{results:rq2} (RQ2). We describe developers' motivations for using unsafe code in \tlink{Section}{results:rq3} (RQ3), and we report how they reason about encapsulation in \tlink{Section}{results:rq4} (RQ4). We summarize key results at the end of each section. 

\subsection{Demographics}
\label{results:demo}

\begin{table}
\caption{Demographics of interview and survey participants.}
\label{figure:results:demo}
\begin{subtable}[t]{0.35\textwidth}
\caption{Interview participants' affiliations ("Affil.") and years of experience using Rust ("Years") by ID.}
\label{figure:method:demo}
\vskip 0pt
\begin{minipage}[t]{0.5\textwidth}
\vskip 0pt
\centering
\tiny
\begin{tabular}{ccc}
\textbf{ID}                        & \textbf{Affil.} & \textbf{Years} \\ \hline
\rowcolor[HTML]{EFEFEF} \textbf{1}  & I                    & 2                        \\
\textbf{2}                          & I,R                  & 10                   \\
\rowcolor[HTML]{EFEFEF} \textbf{3}  & I,O                  & 5                   \\
\textbf{4}                          & I,O                  & 8                  \\
\rowcolor[HTML]{EFEFEF} \textbf{5}  & I,O                  & 2                  \\
\textbf{6}                          & A                    & 3                 \\
\rowcolor[HTML]{EFEFEF} \textbf{7}  & I                    & 6                  \\
\textbf{8}                          & I                    & 5               \\
\rowcolor[HTML]{EFEFEF} \textbf{9}  & O                    & 2              \\
\textbf{10}                         & O                    & 2              \\ \hline
\end{tabular}
\end{minipage}\begin{minipage}[t]{0.5\textwidth}
\vskip 0pt
\tiny
\begin{tabular}{ccc}
\textbf{ID}                        & \textbf{Affil.} & \textbf{Years} \\ \hline
\rowcolor[HTML]{EFEFEF} \textbf{11} & I                    & 5                 \\
\textbf{12}                         & I,A                  & 6                \\
\rowcolor[HTML]{EFEFEF} \textbf{13} & I                    & 10                \\
\textbf{14}                         & R,O,I                & 7               \\
\rowcolor[HTML]{EFEFEF} \textbf{15} & I                    & 1.5            \\
\textbf{16}                         & I                    & 3             \\
\rowcolor[HTML]{EFEFEF} \textbf{17} & A                    & 4             \\
\textbf{18}                         & I                    & 3                     \\
\rowcolor[HTML]{EFEFEF} \textbf{19} & I                    & 4                    \\ \hline
\end{tabular}
\end{minipage}\\
\scriptsize
\centering
\vskip 1em
{
\setlength{\tabcolsep}{1pt}
\begin{tabular}{cclccl}
I &=& Industry &\quad R &=& Rust Team\\
A &=& Academia &\quad O &=& Open Source\\
\end{tabular}
}
\end{subtable}\hfill\begin{subtable}[t]{0.6\textwidth}
\caption{The mean ($\mu$) and standard deviation ($\sigma$) of survey respondents' years of experience in the field of software engineering ("SE"), as well as using Rust, C and C++. Entries are grouped by where respondents indicated that they had first heard about the survey. Respondents could select multiple options.}
\label{figure:results:demo:survey-years}
\centering 
\scriptsize
\begin{tabular}{lccccccccc}
                                                                                                                        & \multicolumn{1}{l}{}     & \multicolumn{2}{c}{\textbf{SE}}           & \multicolumn{2}{c}{\textbf{C}}                            & \multicolumn{2}{c}{\textbf{C++}}                         & \multicolumn{2}{c}{\textbf{Rust}}         \\ \cline{3-10} 
\textbf{Location}                                                                                                       & \textbf{\#}              & $\mu$     & $\sigma$                 & $\mu$ & $\sigma$                 & $\mu$ & $\sigma$               & $\mu$ & $\sigma$ \\ \hline\rowcolor[HTML]{EFEFEF}  Rust Forums                                                                & \multicolumn{1}{c}{16}  & 17.3 & \multicolumn{1}{c|}{10.1} & 7.5                  & \multicolumn{1}{c|}{6.8}  & 5.6                  & \multicolumn{1}{c|}{6.2} & 4.5                  & 2.4       \\
Rust Discord                                                                                                            & \multicolumn{1}{c}{10}  & 8.0  & \multicolumn{1}{c|}{3.7}  & 2.7                  & \multicolumn{1}{c|}{2.8}  & 3.4                  & \multicolumn{1}{c|}{2.3} & 3.6                  & 2.0       \\
\rowcolor[HTML]{EFEFEF}  Reddit (\href{https://www.reddit.com/r/rust/}{/r/rust}) & \multicolumn{1}{c}{83}  & 13.9 & \multicolumn{1}{c|}{9.1}  & 7.6                  & \multicolumn{1}{c|}{7.2}  & 6.0                  & \multicolumn{1}{c|}{6.3} & 4.1                  & 2.2       \\
"This Week in Rust"                                                                                                     & \multicolumn{1}{c}{52}  & 15.1 & \multicolumn{1}{c|}{9.7}  & 6.6                  & \multicolumn{1}{c|}{7.7}  & 5.7                  & \multicolumn{1}{c|}{6.9} & 4.1                  & 2.3       \\
\rowcolor[HTML]{EFEFEF}  Other                                                                      & \multicolumn{1}{c}{3}   & 23.7 & \multicolumn{1}{c|}{15.8} & 17.3                 & \multicolumn{1}{c|}{14.6} & 9.0                  & \multicolumn{1}{c|}{5.6} & 3.7                  & 1.5       \\
All                                                                                                                     & \multicolumn{1}{c}{160} & 14.5 & \multicolumn{1}{c|}{9.5}  & 7.2                  & \multicolumn{1}{c|}{7.5}  & 5.8                  & \multicolumn{1}{c|}{6.3} & 4.1                  & 2.2       \\ \hline
\end{tabular}
\end{subtable}
\end{table}

\tlink{Table}{figure:results:demo} provides summary statistics on interview and survey participants.
We invited all 42 eligible candidates to participate in interviews, and we interviewed 19 who responded to our invitation. \tlink{Table}{figure:method:demo} summarizes interview participants' demographic information. There were 15 who had experience in industry, six in open source development, three in academia, and two were members of the Rust Team. They had 4.7 years of experience with Rust on average, and the most experienced participants had been using Rust for 10 years. We did not collect any additional demographic information from our interview participants. 

We received 368 survey responses, of which 240 were complete, 203 met our eligibility criteria, and 160 passed each of our measures for fraud detection. Hereafter, when we refer to ``respondents,'' we include only the 160 that passed each of our checks. We collected several categories of demographic information from survey respondents. The majority were young, educated, identified as male, and had experience in either industry or open source development. Nearly 66\% had completed at least a bachelor's degree, and 30\% had completed a graduate degree. Most of the respondents (83\%) were under 40 years of age, and 56\% were between the ages of 18 and 29. We followed Spiel et al.'s~\cite{spiel19_gender} guidelines for ethically surveying respondents' gender; 82\% identified as ``Man'', 6\% identified as ``Non-binary'', 3\% identified as ``Woman'' and 9\% did not disclose their gender. \tlink{Table}{figure:results:demo:survey-years} summarizes respondents' years of experience with the field of software engineering, Rust, and other systems languages. On average, respondents had more than a decade of experience in software engineering, more than 5 years of experience with C and \CC{}, and just over 4 years of experience using Rust. 

\begin{table}
\footnotesize
\centering
\caption{How frequently survey respondents ``engaged with'' and wrote unsafe code. Percentages on the left include ``Less than once a year'' and ``Yearly'', while those on the right include ``Weekly'' and ``Daily''. The number of respondents in the sample is listed in parentheses on the righthand side.}
\label{figure:results:engagement}
\begin{center}
\noindent\begin{tblr}{width=\textwidth,colspec={Q[c,m]X[l,m]Q[c,m]Q[c,m]}}
\textbf{ID} & \textbf{Question} & \textbf{Sample} & \textbf{Distribution} \\ \hline
\qlabel{q:demo:freq-engage} & How frequently do you engage with unsafe Rust code in any way? & \SetCell[r=2]{c} All & \tzplotbars{12.5}{28.1}{36.2}{18.1}{5}{160} \\ \hline
 \qlabel{q:demo:freq-write}  & How frequently do you write new Rust code or edit existing Rust code within an unsafe block or function?         &  &          \tzplotbars{7.5}{18.1}{43.1}{20.6}{10.6}{160}  \\ \hline  
\end{tblr}
\end{center}
\cbox{l1} \ \  Less than once a year \quad \cbox{l2} \ \  Yearly \quad \cbox{mid} \ \ Monthly \quad \cbox{r1} \ \ Weekly \quad \cbox{r2} \ \ Daily \\
\end{table}
Participants had a variety of roles within the Rust community and differing levels of engagement with unsafe code. Among survey respondents, 74\% had committed unsafe code to a public GitHub repository, and 66\% had contributed to a published crate. Our interview population had regularly written or edited unsafe code, but 47\% of our survey population did not. Since ``regularly'' is subjective, we asked the questions shown in \tlink{Table}{figure:results:engagement} to provide a more specific time-frame and to distinguish editing unsafe code from practices that do not require code contributions, such as auditing. Only 26\%~(\ref{q:demo:freq-write}) of survey respondents would write or edit unsafe code more than once a month, but 41\%~(\ref{q:demo:freq-engage}) would \textit{engage} with it at this frequency.

\subsection{RQ1 - Interoperation}
\label{results:rq1}
The majority of interview participants and 70\% of survey respondents used unsafe code to call foreign functions. The five most popular languages that respondents interoperated with were C (90\%), C++ (57\%), JavaScript or TypeScript (22\%), Assembly (21\%), and Python (17\%). With the exception of Assembly, the relative popularity of each language is similar to results from Fulton et al.~\cite{fulton21}. Their survey included Assembly as part of the "Other" category, which only 7\% of their respondents had selected. Although 72\% of our respondents had converted part or all of an application written in another language to Rust, 25\% of these respondents had not called foreign functions. This subset might have ported these applications wholesale instead of opting for an incremental approach.

\subsubsection*{Aliasing \& Memory Management} Rust's standard library provides several types that can be used to manage ownership of heap allocations. Each of these types can be converted to raw pointers when necessary, but similar to Rust's safe references, they have distinct invariants that must be maintained to avoid undefined behavior. Participants frequently used these types to allocate memory for foreign function calls, converting them into raw pointers in the process. In particular, 54\% of survey respondents who had called foreign functions also used one of Rust's memory container types for this purpose. Rust's thread-safe \code{Arc}---an atomic, reference-counted allocation---was used by 28\% of these participants, while only 8\% had used the non-atomic \code{Rc}. Rust's \code{Box} provides unique ownership over a heap allocation, and it was used by 95\% of these participants to allocate memory for foreign calls. One interview participant used \code{Box} extensively when porting an online multiplayer text-adventure game from C to Rust. The game's logic was written in C, but all heap memory was allocated in Rust using \code{Box}. When a \code{Box} was unwrapped and exposed as a raw pointer to C, its address was added to a table. The table was checked and updated to prevent double-free errors.

 Several interview participants mentioned having issues with this type, noting that when a \code{Box} is moved, any raw pointers to its contents become invalid. Participants with a background in \CC{} development had not expected \code{Box} to behave this way. They were accustomed to using \code{unique\_ptr} from the \CC{} standard library. This has a similar role, but different semantics:
 \begin{pquote}{9}
 You give the raw pointer off to the interface, and then you relocate the unique pointer, but you still rely on that pointer having valid access tags.
 \end{pquote}
 Survey participants did not encounter this issue as frequently. Of the 51\% of respondents had used \code{Box} in any unsafe context, only 17\% had encountered issues when a pointer became invalid after a \code{Box} was moved. However, we did not determine if respondents had actually unwrapped a \code{Box} or otherwise retained a raw pointer to its contents, so this 17\% could still represent a significant portion of the respondents who would have been able to encounter this problem. Our population is most representative of experienced Rust developers, so it is possible that the average Rust developer is even more likely to have misused Rust's \code{Box} type in these contexts.
 
One participant observed several aliasing patterns in foreign libraries that were at odds with Rust's restrictions. They struggled to encapsulate foreign functions that would read from a ``source'' pointer and write to a ``destination'' pointer. It was difficult to use safe references as parameters for the Rust encapsulations of these functions when the source and destination pointers needed to be aliases. The borrow checker would prevent the mutable destination reference from aliasing with the immutable source reference. Interview participants also noted that C allows unrestricted integer-to-pointer conversion, but Rust does not have an agreed-upon semantics for this behavior~\cite{jung24}. Another participant observed that C libraries would often use on \ilquote{a rat's nest of pointers}{14} to encode self-referential structures, which were difficult to represent in Rust encapsulations.

To manage these differences, participants reviewed the documentation and implementation of foreign libraries to determine their safety requirements. However, one participant found that documentation was frequently missing, so they had to manually reason about these properties.
\begin{pquote}{11}
It's really difficult to figure out a lot of the properties of a given codebase...are you actually mutating something?...if I give you some input data, are you going to retain a pointer to that and keep operating on it later?...it's a very human issue of lack of documentation and lack of correctness within a codebase.
\end{pquote}
This pattern matches two of the aliasing violations observed by McCormack et al.~\cite{mirilli}, where "unique" mutable references were cast into raw pointers and stored into the heap by foreign functions.

The reverse of this situation occurred when participants exposed Rust libraries to languages with fewer safety guarantees. Properties that were implicitly enforced by Rust's type system became documentation in other languages. 
\begin{pquote}{13}
In Rust, you can trivially say, I'm going to return to you a thing, which just borrows my memory, and then you just can't access me while you're using that
...this is not very easily mappable into a C API, or rather you could do it, but then you have to read all this documentation...
\end{pquote}
Some foreign APIs were structured in a way that made memory management easier. One participant found that \ilquote{there [were] no complex lifetimes involved}{11} in their encapsulation of an embedded library. Others observed that the properties they typically associated with a \ilquote{well-crafted C and \CC{} API}{15} made it easier to create a Rust encapsulation. For example, one participant observed that C APIs tend to expose pairs of initialization and cleanup functions for each type, which could be linked into the behavior of the \code{Drop} trait for its Rust encapsulation. This trait provides a cleanup function that is called to free the resources associated with an object when it leaves scope. 

\subsubsection*{Concurrency} Interview participants typically used more sophisticated multithreading patterns in the Rust components of multilanguage applications. One participant used granular locking within the Rust codebase, while the C codebase had a singular global lock. For another participant, the \CC{} component of their application was structured along a single main thread, while the Rust component used multithreading freely---\ilquote{we were throwing threads around all the time}{2}. It was easy for this participant to accidentally call into \CC{} from the wrong Rust thread, but they saw this as less challenging to manage than it would have been to implement the entire application in \CC{}:
\begin{pquote}{2}
It was a validation of a belief...that this would have been impossible to do in pure \CC{}...we had to hit this [problem] for the occasional times we had to call out to C++ from Rust...but if we were only writing \CC{} code... we would have been hitting this constantly.
\end{pquote}
However, Rust's approach to concurrency also made it difficult for one participant to encapsulate a foreign library. In Rust, a type can automatically derive the traits \code{Send} and \code{Sync} if all of its components implement these traits. However this participant observed that in C and C++, a single type can have methods and components with varying thread-safety properties: \ilquote{Rust enforces thread safety based on types... C tends to do it based on functions}{14}. They found that it was more difficult to encapsulate foreign types that were only partially thread-safe.
\begin{table}
\footnotesize
\centering
\caption{How often survey respondents would avoid passing abstract data types by value to foreign functions and converting foreign raw pointers into safe references. Percentages on the left include ``Always'' and ``Most of the time'', while those on the right include ``Sometimes'' and ``Never''. The number of respondents in the sample is listed in parentheses on the righthand side.}
\label{table:ffi:types}
\begin{tblr}{width=\textwidth,colspec={Q[c,m]X[l,m]Q[c,m]Q[c,m]}}
\textbf{ID} &
\textbf{Question} &\textbf{Sample}                                   & \textbf{Distribution} \\ \hline
\qlabel{q:ffi:adt} & How often do you intentionally avoid passing Rust's abstract data types (structs, enums) by value across FFI boundaries? & \SetCell[r=2]{c} Used FFI & \tzplotbarsunsure{4.5}{19.6}{19.6}{7.1}{27.7}{21.4}{112}
\\ \hline
\qlabel{q:ffi:ptr} & How often do you intentionally avoid converting raw pointers to memory allocated by FFI calls into safe references, such as \code{\&T} or \code{\&mut T}? &               & \tzplotbarsunsure{7.1}{35.7}{23.2}{10.7}{12.5}{10.7}{112}              \\ \hline
\SetCell[c=4]{c}\cbox{l1} \ \  Always \quad \cbox{l2.1} \ \  Most of the time \quad \cbox{l2.2} \ \ About half the time \quad \cbox{mid} \ \ Unsure \quad \cbox{r1} \ \ Sometimes \quad \cbox{r2} \ \ Never
\\
\end{tblr}
\end{table}

\subsubsection*{Information-Hiding} Participants attempted to keep interoperation with foreign code as straightforward as possible. When declaring bindings, one participant preferred using primitive types and pointers instead of abstract data types like \code{Option} and \code{NonNull}, even though Rust allows these types to be implicitly cast into raw pointers~\cite{null_ptr_opt}. They felt that these casts hid useful contextual information, so they preferred to use raw pointers explicitly. A few participants indicated that they had avoided passing structs by value across foreign boundaries in certain situations. One of these participants observed that certain Rust programs would behave incorrectly when structs were moved across a foreign boundary instead of copied, or if they had a destructor on the \CC{} side of the boundary. They typically avoided passing structs by value in either situation, and they recommended that \ilquote{...if you're doing FFI, the only things you should pass by value are copy types}{2}. Other interview participants also indicated that they felt comfortable passing structs by value as long as they had copy semantics and equivalent types on each side of the boundary.

In addition to limiting the use of certain types at boundaries, interview participants also attempted to minimize the interactions between each side of the foreign function boundary. They thought that a \ilquote{very chatty API}{7} would be difficult to encapsulate correctly. One participant described this complexity in terms of the heap object graph that a foreign library exposes to Rust. When the structure of the foreign heap was exposed to Rust in detail, safe encapsulations became difficult to use:
\begin{pquote}{14}
It's hard to keep those objects around for very long unless you want all of your structs to end up with a bunch of lifetimes...
you don't want that...
it'll scare away new programmers for sure.
\end{pquote}
They typically avoided casting foreign pointers into Rust's safe reference types since \ilquote{applying lifetime semantics to foreign code}{4} required difficult, manual reasoning. Another participant used opaque types to hide the underlying layout of foreign values. This design pattern is useful in situations where Rust needs to be able to reference a foreign value, but the value does not need to be accessed on the Rust side of the boundary. Instead of declaring the structure of its type on each side of the boundary, a value can be given a zero-size, ``opaque'' placeholder type in Rust. 

We investigated these information-hiding practices using the survey questions shown in \tlink{Table}{table:ffi:types}, where respondents indicated if they ``intentionally avoid'' passing structs by value or casting foreign pointers into Rust's references. We added ``Unsure'' as a response option for these questions since they indirectly rely on participants having a notion of why they would ``intentionally'' avoid these behaviors. Neither practice was consistently followed, but slightly more respondents would avoid passing abstract data types by value most or all of the time (49\%, \ref{q:ffi:adt}), than would avoid casting foreign pointers into references (43\%, \ref{q:ffi:ptr}).
\vskip 0.5\baselineskip
\noindent\textbf{Key Findings - RQ1 (Interoperation).} \\
\rsq{How do Rust developers reason about memory safety across foreign function boundaries?}
\label{rq1}
\vskip 0.5\baselineskip
\noindent The majority of participants called foreign functions, which were typically written in C and C++. Certain foreign libraries used aliasing and concurrency patterns that conflicted with the expectations of Rust's type system, which made them difficult to encapsulate. To make interoperation easier, participants minimized their interaction with foreign libraries, but they frequently used Rust's memory containers as allocators, and few actively avoided casting foreign pointers into safe references. Developers identified safety properties and design patterns of foreign APIs that match the characteristics of libraries where McCormack et al.~\cite{mirilli} detected aliasing violations, indicating that foreign function calls may be a significant unchecked source of undefined behavior in the Rust ecosystem.

\subsection{RQ2 - Tooling}
\label{results:rq2}

\begin{table}
\centering
\footnotesize
\caption{How often survey respondents audited their dependencies, used a debugger, wrote tests, and executed tests in Miri (if they had used Miri at least once). Percentages on the left include ``Always'' and ``Most of the time'', while those on the right include ``Sometimes'' and ``Never''. The number of respondents in each sample is included in parentheses on the righthand side.}
\label{results:rq2:tests-and-auditing}
\begin{tblr}{width=\textwidth,colspec={Q[c,m]X[l,m]Q[c,m]Q[c,m]}}

\textbf{ID} &\textbf{Question} & \textbf{Sample} & \textbf{Distribution} \\ \hline
\SetCell[r=2]{c,m} \qlabel{q:tool:wrote-tests} &  \SetCell[r=2]{l} How often do you write tests for Rust applications that use \code{unsafe}? & All & \tzplotbars{5.6}{24.4}{9.4}{28.7}{31.9}{160}         \\ \hline
 &  & Used Miri    & \tzplotbars{3.1}{16.3}{9.2}{33.7}{37.8}{98}     \\ \hline
                  
\qlabel{q:tool:ran-tests-miri} & How often do you run test cases in Miri? & Used Miri & \tzplotbars{13.3}{42.9}{6.1}{24.5}{13.3}{98}         \\ \hline
\SetCell[r=2]{c,m} \qlabel{q:tool:audit} &  \SetCell[r=2]{l} How often do you audit your dependencies' use of \code{unsafe}? & All & \tzplotbars{48.1}{40}{5.6}{4.4}{1.9}{160}         \\ \hline
& & Used Auditing Tools  & \tzplotbars{25.9}{54.3}{9.9}{6.2}{3.7}{81}       \\ \hline
\SetCell[c=4]{c}\cbox{l1} \ \  Always \quad \cbox{l2} \ \  Most of the time \quad \cbox{mid} \ \ About half the time \quad \cbox{r1} \ \ Sometimes \quad \cbox{r2} \ \ Never
\end{tblr}
\end{table}

\begin{table}
\centering
\footnotesize
\caption{Whether or not survey respondents who had used foreign functions would trust the correctness of foreign function bindings written by hand and generated using a tool. Percentages on the left include ``Definitely yes'' and ``Probably yes'', while those on the right include ``Probably not'' and ``Definitely not''. The number of respondents the each sample is included in parentheses on the righthand side.}
\label{results:rq2:bindings}
\begin{tblr}{width=\textwidth,colspec={Q[c,m]X[l,m]Q[c,m]Q[c,m]}}
\textbf{ID} & \textbf{Question}    & \textbf{Sample}                     & \textbf{Distribution}               \\ \hline
\qlabel{q:tool:bindings-hw} & Do you trust FFI bindings that are written by hand? & \SetCell[r=2]{m} Used FFI &\tzplotbars{2.7}{4.5}{48.2}{35.7}{8.9}{112} \\ \hline
\qlabel{q:tool:bindings-gen} & Do you trust FFI bindings that are generated by a tool?           &  & \tzplotbars{0}{1.8}{19.6}{60.7}{17.9}{112} \\ \hline
\SetCell[c=4]{m}\cbox{l1} \ \  Definitely yes \quad \cbox{l2} \ \  Probably yes \quad \cbox{mid} \ \ Might or might not \quad \cbox{r1} \ \ Probably not \quad \cbox{r2} \ \ Definitely not
\end{tblr}
\end{table}

Participants used a wide variety of development tools to assist with validating their design choices. We focused primarily on dynamic bug-finding tools and static tools for generating bindings to foreign functions. However, certain participants also described their experiences using debuggers, verifiers, and auditing tools. 

\paragraph{Dynamic Tools} Most participants used dynamic bug finding tools. Interview participants often mentioned Miri~\cite{miri}, AddressSanitizer~\cite{asan}, and Valgrind~\cite{seward07}. We prompted survey respondents to indicate which dynamic tools they had used from a list of options mentioned by our interview participants. Survey respondents most frequently used Miri (61\%), Valgrind (44\%), AddressSanitizer (26\%), and \code{cargo-fuzz}~\cite{cargofuzz} (26\%). Each of the remaining tools listed in our survey was used by less than 15\% of respondents. Three interview participants and 31\% of survey respondents used some form of fuzzing tools. One interview participant leveraged industry sponsorship to fuzz a JIT compiler for hours-on-end. Another participant implemented a BTree data structure with optimizations for high performance computing, and they used \code{libfuzzer}~\cite{libfuzzer} to perform differential testing~\cite{mckeeman98} against the implementation provided by Rust's standard library. Of the 26 developers interviewed by Höltervennhoff et al.~\cite{holtervennhoff23} only seven had used Miri, seven used fuzzing tools, and three used Valgrind. Our results indicate that this is not representative of a broader trend in the Rust community. 

The majority of survey respondents also wrote tests for applications that used unsafe code.  As shown in \tlink{Table}{results:rq2:tests-and-auditing}, 61\%~(\ref{q:tool:wrote-tests}) of all participants wrote tests most of the time or always. This was equally true for the subset of participants who had used Miri; 72\% would write test cases at least most of the time. However, 56\%~(\ref{q:tool:ran-tests-miri}) of survey respondents who had used Miri only sometimes used it to run their test cases, if at all. Höltervennhoff et al.~\cite{holtervennhoff23} also had similar findings. The majority of their participants wrote unit tests for unsafe code, but a subset claimed that they would infrequently run these tests because foreign function calls or low-level hardware interaction made testing ``impossible.'' While notable, we cannot make any causal claims between Miri's limitations and this lack of testing in our survey population. Additionally, we did not ask if survey respondents regularly ran their test cases. However, interview participants frequently mentioned that Miri's slow performance and lack of support for foreign function calls prevented them from using it to find bugs in certain categories of applications, and survey respondents who had used Miri encountered similar issues. The majority (62\%) were deterred from using it due to one or more of the following issues: lack of support for foreign functions (43\%), slow performance (26\%), or lack of support for inline assembly (19\%). Developers interviewed by Höltervennhoff et al.~\cite{holtervennhoff23} had also struggled with Miri's lack of support for key features, but none had mentioned its performance.

Notably, one of our interview participant had neglected to use Miri since their application relied heavily on foreign function calls: \ilquote{we've assumed that we are a JIT and Miri can't run JIT code, so therefore it's useless}{13}. However, after a vulnerability was discovered in their codebase, they decided it that it would be worthwhile to run Miri on the subset of their test cases that did not require foreign function calls. They discovered that Miri would have been able to detect the vulnerability earlier, since it was caused by a violation of Rust's aliasing model. Another interview participant managed to circumvent Miri's limitations by creating mocks of foreign functions, but this solution did not scale to programs with \ilquote{50,000 lines of code}{4}. This participant and a few others were aware of the Krabcake project~\cite{krabcake}, which aims to extend Valgrind with Miri's borrow tracking features. Each of these participants was already using Valgrind with Rust codebases, and they felt that Krabcake would be an ideal solution for overcoming Miri's limitations.

\paragraph{Foreign Function Bindings} Interview participants encountered several issues with tools that generate bindings to foreign functions. Participants reported that \code{bindgen}~\cite{bindgen} added significant runtime overhead during build times and was not as expressive as they desired. One participant observed that \code{bindgen} \ilquote{did some weird stuff for anonymous unions and structs}{17} but that this was easy to fix by modifying the tool's configuration. However, another participant found it necessary to switch from using \code{bindgen} to writing bindings by hand, because it was not capable of expanding the macros that they had used to define core data structures. Others felt that the bindings they wrote by hand were more idiomatic than the output of a tool. The majority (58\%) of survey respondents who created bindings used multiple methods, while 28\% only generated bindings and 14\% only wrote them manually.

Interview participants who wrote bindings by hand viewed this process as error-prone, noting that bindings become a \ilquote{fiddly mess}{5} when you need to ensure that interfaces remain consistent on each side of the foreign boundary. These errors clearly do occur in practice, as 35\% of survey respondents who had called foreign function had also encountered incorrect bindings. McCormack et al.~\cite{mirilli} also found several instances of incorrect bindings in Rust libraries. Some were intentional\textemdash developers neglected to add return types for values they did not use\textemdash but others introduced aliasing violations or uninitialized reads. Survey respondents who had called foreign functions were somewhat more likely to trust generated bindings. As shown in \tlink{Table}{results:rq2:bindings}, only 45\%~(\ref{q:tool:bindings-hw}) of respondents would at least ``probably'' trust handwritten bindings, compared to 79\%~(\ref{q:tool:bindings-gen}) who would trust generated bindings. However, developers were still more likely to trust handwritten bindings than to distrust them.

\paragraph{Debugging} Few interview participants reported using a debugger with Rust code. Only 10\% of survey respondents used a debugger daily, 29\% used one monthly, and 44\% used one yearly at most, if at all. One interview participant had used both MSVC's debugger and LLDB~\cite{lldb}, and while both worked, neither matched the quality of the other parts of the Rust toolchain, \ilquote{where you get all the bells and whistles that you want, and things just work nicely together}{3}. Mozilla's RR debugger~\cite{rr_debugger} was useful for one participant to resolve bugs found from fuzzing, but it was generally more common for participants to report using \ilquote{\texttt{printf}-style}{7} debugging. 

\paragraph{Formal Methods} Few interview participants and only 10 survey respondents used tools that apply formal methods for static and dynamic verification. Interview participants mentioned using Kani~\cite{kani}, Prusti~\cite{prusti_astrauskas19}, and Crucible~\cite{crucible}. Kani was useful for one participant, who needed to prove that a particular routine would never execute more than twice for any input. Another participant who had \ilquotenonum{tried a bunch of the tools} in this category found that Prusti was the most complete, but they felt that it would be more effective if these tools were \ilquotenonum{inherently parts of Rust} so that they could \ilquote{hoist a lot of currently unsafe code into safe code}{11}. Of the ten survey respondents who had used ``formal methods tools,'' five had used Kani, three had used Prusti, three had used Creusot~\cite{creusot_denis22}, one had used Flux~\cite{lehmann23}, and one had used Crucible. We provided a short-response option for this question to include any tools that were missing from our list. 

\paragraph{Auditing} Only a few interview participants indicated that they had ever audited their dependencies. Höltervennhoff et al.~\cite{holtervennhoff23} had similar findings; half of the developers that they interviewed had deliberately included dependencies with unsafe code, but the majority did not audit their dependencies. One of our participants would examine their dependencies if they were curious about how they functioned, and they actively avoided adding dependencies that used outdated versions of Rust or deprecated features. Another participant was in the process of having a third-party institution examine their codebase to determine if they could reduce their amount of unsafe code. 

Survey respondents rarely audited their dependencies' use of unsafe code. More than 80\%~(\ref{q:tool:audit}) of respondents only sometimes audited their dependencies' use of unsafe code, if at all. It is possible that a subset of the participants who answered ``Never'' to this question did not have any dependencies to audit. However, 51\% of our respondents had used automated auditing tools, compared to four interviewed by Höltervennhoff et al.~\cite{holtervennhoff23}. Of this subset, 63\% had used \code{cargo-audit}~\cite{cargoaudit}, 42\% had used \code{cargo-update}~\cite{cargoupdate}, 31\% had used \code{cargo-deny}~\cite{cargodeny}, 22\% had used \code{cargo-geiger}~\cite{cargogeiger}, and only 7\% had used \code{cargo-vet}~\cite{cargovet}.

\vskip \baselineskip
\noindent\textbf{Key Findings - RQ2 (Tooling).} \\
\noindent\rsq{What tools do Rust developers use when contributing to applications that include unsafe code, and how could tooling be improved?}
\label{rq2}
\vskip 0.5\baselineskip
\noindent Miri appears to be the de facto bug finding tool for the Rust ecosystem, since it was used by 61\% of all respondents. However, 62\% of the respondents who had used Miri before were deterred from using it again due to its performance and lack of support for key features---predominately, foreign function calls. Tools for generating foreign function bindings were often slow and required additional configuration, so a subset of developers still preferred writing bindings by hand. Participants infrequently used debuggers, and only 10 survey respondents had used formal methods tools. Survey respondents infrequently audited their dependencies, but 51\% had used one or more auditing tools. Improvements to Miri may be the most effective way to provide stronger safety guarantees for the majority of our audience, since it was the most widely used tool in our survey. 
\subsection{RQ3 - Motivations}
\label{results:rq3}
\begin{table}
\caption{Survey respondents' motivations for using unsafe code, as measured by the question shown in \tlink{Figure}{figure:method:questions}. All but one participant chose to answer this question. Each cell shows the percentage of participants who identified with both of the motivations for that row and column. All three motivations were selected by 7\% of participants.}
\label{table:motivations}
\footnotesize
\begin{tabular}{rccc}
\multicolumn{1}{r|}{}                   & \textbf{Necessity}           & \textbf{Performance}         & \textbf{Ergonomics}                               \\ \hline
\multicolumn{1}{r|}{\textbf{Necessity}} & \cellcolor[HTML]{EFEFEF}77\% & \cellcolor[HTML]{FFFFFF}92\% & \multicolumn{1}{c|}{\cellcolor[HTML]{EFEFEF}86\%} \\ \cline{2-2}
\textbf{Performance}                    & \multicolumn{1}{c|}{}        & \cellcolor[HTML]{EFEFEF}47\% & \multicolumn{1}{c|}{\cellcolor[HTML]{FFFFFF}53\%} \\ \cline{3-3}
\textbf{Ergonomics}                     &                              & \multicolumn{1}{c|}{}        & \multicolumn{1}{c|}{\cellcolor[HTML]{EFEFEF}18\%} \\ \cline{4-4} 
\end{tabular}
\end{table}

Participants used unsafe code because they perceived that it was more performant or ergonomic than safe alternatives, or that there were no safe alternatives at all. Here, we refer to these motivations as ``Performance,'' ``Ergonomics,'' and ``Necessity,'' which correspond to the codes ``Increase Performance'', ``Easier or More Ergonomic,'' and ``No Other Choice'' from \tlink{Figure}{figure:method:codebook}, respectively. 
We used the question shown in \tlink{Figure}{figure:method:questions} to measure the distribution of each of these motivations in our survey population. All but one participant chose to answer this question. Results are shown in \tlink{Table}{table:motivations}. Necessity was the most common motivation, followed by performance and ergonomics. 

\begin{table}
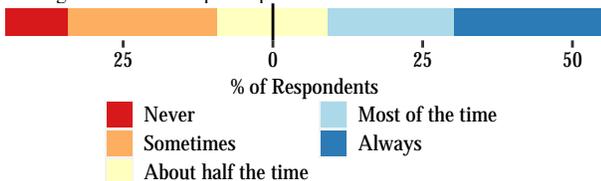
\caption{How frequently participants measured the impact of their changes when using unsafe code to improve performance. Percentages on the left include ``Always'' and ``Most of the time'', while those on the right include ``Sometimes'' and ``Never''. The number of respondents in the sample is included in parentheses on the righthand side.}
\label{figure:motivations:performance}
\footnotesize
\begin{tblr}{width=\textwidth,colspec={Q[c,m]X[4,l,m]X[1,c,m]Q[c,m]}}
\textbf{ID} &
\textbf{Question}           & \textbf{Sample}                                                                            & \textbf{Distribution}              \\ \hline
\qlabel{motivation:profiling} & 
 When you choose to use \code{unsafe} because it performs faster or is more space efficient, how often do you measure the difference? & Used \code{unsafe} for performance & \tzplotbars{10.5}{25}{18.4}{21.1}{25}{76} \\ \hline
\SetCell[c=4]{c} \SetCell[c=4]{m}\cbox{l1} \ \  Always \quad \cbox{l2} \ \  Most of the time \quad \cbox{mid} \ \ About half the time \quad \cbox{r1} \ \ Sometimes \quad \cbox{r2} \ \ Never
\end{tblr}
\end{table}

\subsubsection*{Performance}
Some interview participants and 47\% of survey respondents indicated that they would use unsafe code when they perceived that it was faster or more space-efficient than an equivalent safe API. Participants usually identified a safe alternative in these situations, but they felt that it would be too expensive to use: \ilquote{so you could do it, right? But that would be some performance overhead}{12}. Most interview participants who attempted to use unsafe to improve performance had implemented small-scale optimizations in existing applications. These typically involved eliminating runtime checks, which were seen as unnecessary due to what they perceived were local or global invariants. For example, one participant used an unsafe API to consume a pointer to a string without checking if it contained valid UTF-8 characters. This pointer was provided by a foreign call to Python, and the participant thought that it would only ever provide text in a valid format. Another participant chose to store a heap allocation as a raw pointer in a static, mutable variable instead of using one of Rust's safe static encapsulations, since they felt confident that the allocation would remain valid for the duration of the program.

However, a few interview participants used unsafe code to create entire components that were purpose-built to achieve performance. In each situation, performance was seen as a functional requirement for the application domain: \ilquote{I write a serialization framework...and obviously a goal for it is to be extremely performant}{3}. A participant who took this approach found that it significantly increased the amount of unsafe code in their application, but they felt that this was a reasonable compromise to achieve greater performance. Survey respondents who reported using unsafe code to improve performance did not have a strong tendency toward either large-scale or small-scale use. We found that 26\% only pursued ``small-scale-optimizations in existing applications,'' 20\% only built new, ``large-scale components,'' and 42\% used unsafe code for performance in either situation.

Participants did not consistently measure the performance impact of unsafe code. In some situations, they relied on their intuition to determine which design decisions would be best: \ilquote{I'll totally admit that...I think this is going to be a hot thing, so I'm going to kind of prematurely optimize}{11}. Survey respondents who used unsafe code to increase performance were also inconsistent about measuring its impact. As shown in \tlink{Table}{figure:motivations:performance}, 46\% of respondents reported measuring performance at least most of the time, while 36\% only sometimes measured performance, if at all. In contrast, Höltervennhoff et al.~\cite{holtervennhoff23} found that 6 participants would only use unsafe code to improve performance when the change was ``noticeable.'' We did not make a distinction between profiling and subjective measures of performance in our survey. However, one of our interview participants performed sophisticated profiling on three versions of a component, each of which had varying amounts of unsafe. The most unsafe-heavy version performed the best, but they opted for one with a moderate amount of unsafe code, balancing safety and performance concerns.

\subsubsection*{Ergonomics}
Participants also chose to use unsafe code when they felt it would be easier or more ergonomic than using a safe API. This was the least common motivation cited by interview participants, and only 18\% of survey respondents identified with this motivation. Höltervennhoff et al.~\cite{holtervennhoff23} indicate that some of their participants would use unsafe code if it ``saved them effort'', but it was unclear if this motivation was typical. Our participants usually stated or implied a choice between using safe or unsafe design patterns. For example, one participant chose to use only Rust's unsafe allocator API rather than a mix of unsafe and safe allocation operations. They found that this approach was easier to reason about, since \ilquote{if there's too much abstraction, you lose the information that you need to make it actually sound}{4}. Several participants used the unsafe function \code{transmute} to perform unrestricted type conversion. Since types have differing invariants, \code{transmute} can only be used in unsafe contexts. One participant used transmutation to simplify the implementation of chess engine, which used three enumerations to represent the position of a chess piece. They found that it was easier to use integer conversion, bitwise operations, and transmutation to convert between these enumerations instead of safely matching on their values. Another participant used transmutation when working with a Rust encapsulation of a C font library. This library exposed C heap allocations to Rust as references with short lifetimes\textemdash even though these objects would remain valid on the heap for the duration of the program. This participant found it was easier to transmute the Rust objects to have the \code{{\textquotesingle}static} lifetime, which is indefinite, instead of adapting to the restrictions of the encapsulation. 
\begin{pquote}{14}
I just...transmuted that one to {\normalfont \code{{\textquotesingle}static}}...
another use case for unsafe is working around a bad API when you know that you can use it correctly.
\end{pquote}

\begin{table}
\caption{How frequently participants who felt that they had ``no clear alternative'' to using unsafe code were certain it would be impossible to avoid it. The percentage on the left includes ``Always'' and ``Most of the time'', while the one the right includes ``Sometimes'' and ``Never''. The number of respondents in the sample is included in parentheses on the righthand side.}
\label{figure:motivations:necessity}
\footnotesize
\begin{tblr}{width=\textwidth,colspec={Q[c,m]X[4,l,m]X[1,c,m]Q[c,m]}}
\textbf{ID} & \textbf{Question}           & \textbf{Sample}                                                                            & \textbf{Distribution}              \\ \hline
\qlabel{motivation:necessity} & When you feel that you have no clear alternative other than using unsafe, how often are you certain that it would be completely impossible to accomplish this task using a safe design pattern? & Had no alternative to \code{unsafe} code
 & \tzplotbarsunsure{1.6}{15.4}{3.3}{14.6}{51.2}{13.8}{123} \\ \hline
\SetCell[c=4]{c} \cbox{l1} \ \  Always \quad \cbox{l2.1} \ \  Most of the time \quad \cbox{l2.2} \ \ About half the time \quad \cbox{mid} \ \ Unsure \quad \cbox{r1} \ \ Sometimes \quad \cbox{r2} \ \ Never
\\
\end{tblr}
\end{table}

\subsubsection*{Necessity}

In most situations, participants perceived that they had ``no clear alternative'' to using unsafe code. The majority of interview participants and 77\% of survey respondents identified with this motivation. Höltervennhoff et al.~\cite{holtervennhoff23} observed this to a similar extent; 18 out of their 26 participants claimed to use unsafe code by necessity. Our interview participants often reported being constrained by a combination of Rust's type system and the properties of their applications. For example, one participant found that it was necessary to implement \code{Send} and \code{Sync} in a single-threaded application to maintain compatibility with an existing API. They perceived that the implementation was sound by construction. However, both of these traits are unsafe and only be implemented in an unsafe context.

Participants who contributed to just-in-time (JIT) compilers and operating systems found it difficult to avoid accessing memory through raw pointers. One participant found that this was necessary to implement an unwinding table: \ilquote{I have to jump to that address and I really can't do anything to verify it...}{12}. In other situations, developers felt that there could be a safe alternative, but they were unaware of how to implement it: \ilquote{there's probably a better way that I just don't know about yet or didn't know about at the time and I haven't thought about it}{18}. However, as shown in \tlink{Table}{figure:motivations:necessity}, 65\% of survey respondents were certain at least most of the time, if not always, that it would be impossible to implement an equivalent safe pattern.

\subsubsection*{Co-occurrence} These motivations were not mutually exclusive. One participant implemented a zero-copy deserialization pattern using unsafe code, and they could relate to each of the reasons we identified. Performance was a domain-specific constraint of the internationalization library that they contributing to, which motivated them to use zero-copy deserialization. We reviewed the library's documentation, and it indicated that contributors had considered using an existing crate, but its API was not ergonomic for their use case. The participant perceived that unsafe code was inherently necessary to be able to implement their own version.

\vskip \baselineskip
\noindent\begin{minipage}{\textwidth}
\noindent\textbf{Key Findings - RQ3 (Motivations).} \\
\rsq{What are Rust developers’ motivations for using unsafe code?}
\label{rq3}
\vskip 0.5\baselineskip
Participants were most often motivated to
use unsafe code because they felt that there was no safe
alternative. In other situations, participants used unsafe code
because it performed better or was more ergonomic. For
performance, developers pursued both large-scale abstractions
and small-scale optimizations, but they did not consistently
measure the impact of their changes. Each of these motivations
can influence a design decision simultaneously.
\end{minipage}
\subsection{RQ4 - Encapsulation}
\label{results:rq4}

\begin{table}
\caption{Survey questions relevant to RQ4. Percentages on the left include ``Always'' and ``Most of the time'', while those on the right include ``Sometimes'' and ``Never''. The number of respondents in each sample is included in parentheses on the righthand side.}
\label{results:rq4:frequency}
\footnotesize
\centering
\begin{tblr}{width=\textwidth,colspec={Q[c,m]X[4,l,m]X[1,c,m]Q[c,m]}}
\textbf{ID} & \textbf{Question}           & \textbf{Sample}                                                                             & \textbf{Distribution}              \\ \hline
\qlabel{encap:use-unsafe-api:docs} & When you use an \code{unsafe} API, how often do you look for documentation to ensure that you meet all of its requirements for safety and correctness?          & Used \code{unsafe} APIs & \tzplotbars{1.3}{6.6}{7.2}{22.4}{62.5}{152} \\ \hline

\qlabel{encap:safe-api:checks} & When you expose a safe API for \code{unsafe} code, how often do you include runtime checks to ensure that its requirements for correctness and safety are met? & Exposed a safe API for \code{unsafe} code & \tzplotbars{7.8}{21.3}{12.8}{33.3}{24.8}{141} \\ \hline
\qlabel{encap:unsafe-api:docs} & When you expose an \code{unsafe} API to users, and it is their responsibility to ensure that certain requirements are met, how often do you document these requirements?        & \SetCell[r=2]{c} Exposed an \code{unsafe} API & \tzplotbars{0}{2.9}{7.4}{11.8}{77.9}{68} \\ \hline
\qlabel{encap:unsafe-api:checks} & When you expose an \code{unsafe} API to other users, and it is their responsibility to ensure that certain requirements are met, how often do you include runtime checks for these requirements?     & & \tzplotbars{16.2}{54.4}{11.8}{14.7}{2.9}{68} \\ \hline
\SetCell[r=2]{c,m} \qlabel{encap:refactor} & \SetCell[r=2]{l,m} How often do you refactor Rust applications to remove \code{unsafe} code?                  &  All & \tzplotbars{25}{63.1}{6.2}{5}{0.6}{160} \\ \hline
 &                  &  Regularly wrote \code{unsafe} code & \tzplotbars{15.3}{69.4}{8.2}{7.1}{0}{85} \\ \hline
          \SetCell[r=2]{c,m} \qlabel{encap:avoid} &           \SetCell[r=2]{l,m} How often do you choose to avoid using an \code{unsafe} API when a safe alternative exists?  & All  & \tzplotbars{1.2}{8.8}{7.5}{61.3}{21.2}{160} \\ \hline
               &  & Used \code{unsafe} APIs & \tzplotbars{1.3}{9.2}{6.6}{60.5}{22.4}{152} \\ \hline
\SetCell[c=4]{c}\SetCell[c=4]{c} \SetCell[c=4]{m}\cbox{l1} \ \  Always \quad \cbox{l2} \ \  Most of the time \quad \cbox{mid} \ \ About half the time \quad \cbox{r1} \ \ Sometimes \quad \cbox{r2} \ \ Never
\end{tblr}
\end{table}

\subsubsection*{Invariants}
Participants reasoned about the soundness and correctness of their unsafe code in terms of invariants. When developing operating systems and embedded applications, these invariants were provided by hardware specifications. For example, one participant who developed an embedded application would write a \ilquote{magic value}{8} to RAM to indicate which operation had triggered a reboot. The microcontroller that they used did not erase its memory on reboot, so they could rely on the value being initialized. Some participants perceived hardware-level correctness properties as \ilquote{orthogonal to Rust}{1}, since they were related to the properties of the language. However, other participants who developed embedded applications relied on third-party libraries to encode hardware requirements into the type system so that their programs would fail to compile if a device had been misconfigured.

Developers leveraged Rust's type system to uphold invariants that they believed were necessary for soundness. In operating systems, these typically began as operation-specific refinement properties on primitive types, but they quickly built up into system-level invariants; \ilquote{it gets to a high-level really quickly}{3}. Rust's aliasing restrictions were a natural fit:

\begin{pquote}{12}
We only allow... someone who has access to one of these mapped memory regions to borrow it... That's a great example of taking something that's unsafe inherently and then kind of wrapping it up in a safe abstraction, using the power of Rust to do the borrow checking for us.
\end{pquote}

Another participant experimented with using ghost permissions to reason about pointer arithmetic. This method was popularized for Rust by Yanovski et al.'s \code{GhostCell} type~\cite{ghostcell21}. Other participants relied on ad hoc reasoning to ensure that their programs were correct and free of undefined behavior. Some appealed to their prior experience: \ilquote{{\normalfont [}in{\normalfont ]} \CC{}, you do this all the time}{15}. One participant was accustomed to reasoning about aliasing rules in compiler development, so they felt confident that they were adhering to Rust's restrictions in unsafe contexts. Other participants validated their decisions by auditing their code, but most had some level of implicit trust; \ilquote{that's sort of another... I got to trust it type thing}{5}. This aligns with {Höltervennhoff et al.~\cite{holtervennhoff23}\textemdash a majority of their participants reasoned about unsafe code in terms of contracts or invariants, but some reported using it carelessly, and most felt that it was difficult to write correctly due to its context-sensitivity.

\subsubsection*{Isolation}
The majority of interview participants made a conscious effort to minimize and isolate unsafe code, which made it easier to document and reason about.
\begin{pquote}{10}
That's generally the most important thing, just being self-contained, being well-isolated, being well-encapsulated...
\end{pquote}
Most survey respondents (61\%) predicted that it would be at least somewhat easy for another developer at their skill level to understand a random unsafe block or function from their code. Interview participants who isolated their unsafe code had more confidence that its requirements were met, especially if they went beyond what Rust's type system could verify. However, participants reported that unsafe code was prevalent in JIT compilers and operating systems. Domain-specific, non-local reasoning was necessary to understand any arbitrary \code{unsafe} snippet: \ilquote{you have to know the innards of the JIT-compiled function...so it's just not encapsulated}{11}. Though unsafe code may have been isolated, it is unclear if participants used it minimally. As shown in \tlink{Table}{results:rq4:frequency}, more than 80\% (\ref{encap:refactor}) of survey respondents only sometimes refactored their code to remove unsafe features, if at all. This distribution was similar for the subset of respondents who had regularly written unsafe code.

\subsubsection*{Safe Interfaces}
Most interview participants and 88\% of survey respondents reported that they had exposed a safe API for unsafe code. This was also common for Höltervennhoff et al.~\cite{holtervennhoff23}, who found that 22 of their 26 participants mentioned safe interfacing. Our interview participants would create encapsulations where they perceived that the requirements for their unsafe code were satisfied by either Rust's type system or what they reasoned to be local invariants. 

However 43\% of our survey respondents reported exposing unsafe APIs, relative to six of Höltervenhoff et al.~\cite{holtervennhoff23}'s 26 interview participants. Our participants cited the same motivations for exposing unsafe APIs as they did for using unsafe code in any capacity. One participant exposed unsafe versions of safe API endpoints which had the same behavior, but they did not include runtime checks. Another interview participant exposed unsafe APIs so that users could circumvent their safe encapsulations, in case they became difficult to use. Each of these practices aligns with another participant's perception of a cultural value among Rust developers to make API surfaces as detailed as possible: \ilquote{Rust has a tendency to... make the API as complicated as it needs to be to fully represent what is actually happening behind the scenes}{1}. Other participants indicated that they would only expose an unsafe API when it would be impossible to encapsulate without placing the burden of correctness on the user. Our survey respondents also identified with these motivations; 76\% exposed unsafe APIs because it would be impossible to encapsulate them without preconditions, while 51\% exposed unsafe APIs for performance and 13\% did so for ergonomics. Höltervennhoff et al.~\cite{holtervennhoff23} also found that participants would expose unsafe APIs by necessity or to improve performance, but it was unclear if ergonomics was also a motivation.

Regardless of motivation, 61\% of respondents claimed that they would only expose unsafe APIs when they had requirements beyond what Rust's type system could verify, and they typically documented these requirements. Results for Question~\ref{encap:unsafe-api:docs} indicate that 90\% of respondents who exposed unsafe APIs would document their safety requirements most of the time, if not always. However, respondents did not typically add any additional safeguards to prevent unsafe APIs from being used incorrectly, since 71\%~(\ref{encap:unsafe-api:checks}) only sometimes inserted run-time checks within these APIs, if at all. This was more common when exposing safe APIs; 58\%~(\ref{encap:safe-api:checks}) inserted checks at least most of the time when exposing a safe API for unsafe code. When using unsafe APIs, 85\%~(\ref{encap:use-unsafe-api:docs}) of participants would usually look for documentation to inform their design decisions. 

Interview participants perceived that the Rust community has a strong preference for safety, which affects tool design and development practices. Participants avoided exposing unsafe APIs in libraries partly because of the perception that other Rust developers would otherwise avoid using their crate: \ilquote{everyone wants a safe interface... no one really wants to go use the unsafe heavy ones}{9}. One library author viewed safe encapsulation as an inevitable requirement. Even though exposing an unsafe trait places the burden of responsibility on users, they would still blame the library.

\begin{pquote}{7}
So in theory, the burden of proof is on them, but also if they implement something improperly, then they're going to create a ticket because they'll be like, oh, ``I use your library and [this behavior is] a fault.''
\end{pquote}
An overwhelming majority of survey respondents also preferred safe APIs. More than 80\%~(\ref{encap:avoid}) of respondents claimed that they would choose a safe API over an unsafe alternative most of the time, if not always---regardless of whether or not they had ever used an unsafe API before.

\begin{table}
\caption{Survey questions relevant to RQ4. Percentages on the left include ``Always'' and ``Most of the time'', while those on the right include ``Sometimes'' and ``Never''. The number of respondents in each sample is included in parentheses on the righthand side.}
\label{results:rq4:frequency-unsure}
\footnotesize
\centering
\begin{tblr}{width=\textwidth,colspec={Q[c,m]X[4,l,m]X[1,c,m]Q[c,m]}}
\textbf{ID} & \textbf{Question}           & \textbf{Sample}                                                                             & \textbf{Distribution}              \\ \hline
 \qlabel{q:uncertain:safe-encap} & When you expose a safe API for unsafe code, how often are all of its requirements for correctness and safety satisfied by the properties of Rust's type system? & Exposed a safe API for \code{unsafe} code & \tzplotbarsunsure{0.7}{17.7}{7.8}{11.3}{39}{23.4}{141} \\ \hline
 \qlabel{q:uncertain:docs} & How often is the Rust community's guidance and documentation adequate for you to know how to use unsafe correctly? & All & \tzplotbarsunsure{0.6}{15}{7.5}{6.9}{63.7}{6.2}{160} \\ \hline
\SetCell[c=4]{c}\cbox{l1} \ \  Always \quad \cbox{l2.1} \ \  Most of the time \quad \cbox{l2.2} \ \ About half the time \quad \cbox{mid} \ \ Unsure \quad \cbox{r1} \ \ Sometimes \quad \cbox{r2} \ \ Never
\\ 
 \end{tblr}
\end{table}

\subsubsection*{Uncertainty}
Many interview participants who created safe APIs for unsafe code were uncertain if their encapsulations were sound in all possible situations.
\begin{pquote}{9}
No library author knows when they're going to trigger [undefined behavior]... we do our best to give you a sound interface, but god knows if there's a hole in it...
\end{pquote}
Höltervennhoff et al.~\cite{holtervennhoff23} had similar findings; 16 of their 26 participants felt some degree of uncertainty about whether their unsafe code was correct. Our interview participants were most often uncertain when unsafe was pervasive or when they called foreign functions. When describing their uncertainty about encapsulation, our participants compared the ``theory'' of safety to the ``practice'' of how an API would be used. When these concepts were misaligned, errors occurred. One participant observed this mismatch with a Rust crate that encapsulated a foreign library. The library used a memory-mapped file, and the API made it possible to unmap the file while the user retained a reference to it, leading to a use-after-free error. Another participant struggled with discrepancies between the interface and the implementation of Rust's \code{Layout} API, which describes the size and alignment of an allocation. The design of \code{Layout} was changed to \ilquote{tweak the safety requirements}{14}, which caused code that was previously correct to exhibit undefined behavior. 

Interview participants also had a collective sense of uncertainty about Rust's semantics. They perceived that Rust lacked a formal specification and that the guidelines for unsafe code were incomplete.
\begin{pquote}{10}
I think the biggest issue with Rust's {\normalfont \code{unsafe}} isn't necessarily the tools, but just the spec\textemdash or, rather, the absence of the spec. If you write unsafe code today, you write it against the void.
\end{pquote}
 A few interview participants encountered situations where the precise definition of undefined behavior was either unclear or overly restrictive. For example, one participant felt that it would be necessary to create a mutable reference to uninitialized memory. This is considered undefined behavior by Rust's current semantics~\cite{ucg_ref_ub}, but the participant was uncertain about how this pattern would be problematic in practice. Another participant wished to be able to take a reference to the first field of a product type and use it to create valid references to adjacent fields using integer offsets. This was considered undefined behavior under Rust's Stacked Borrows~\cite{stackedborrows} aliasing model, but it is accepted under the newer Tree Borrows model~\cite{treeborrows}. Additionally, multiple participants were uncertain about how to handle the case when a program panics while a value is being dropped.

Survey respondents were somewhat more certain about their decisions. As shown in \tlink{Table}{results:rq4:frequency-unsure}, more than 60\%~(\ref{q:uncertain:safe-encap}) of respondents who had exposed a safe API for unsafe code were certain at least most of the time, if not always, that its safety properties were satisfied by Rust's type system. Additionally, 70\%~(\ref{q:uncertain:docs}) of respondents found that the Rust community's documentation and guidance was adequate for them to know how to use unsafe code correctly at least most of the time. However, only 6\% felt that it was always adequate, suggesting that improved documentation would be helpful for the overwhelming majority of respondents. Similar to the questions in Tables~\ref{results:rq2:bindings} and ~\ref{figure:motivations:necessity}, we added ``Unsure'' as a response option for the questions shown in \tlink{Table}{results:rq4:frequency-unsure}, since they rely on developers' subjective notions of certainty and adequacy. 

\vskip \baselineskip
\noindent\textbf{Key Findings - RQ4 (Encapsulation).} \\
\noindent\rsq{How do Rust developers reason about encapsulating unsafe code?}
\label{rq4}
\vskip 0.5\baselineskip
\noindent Interview participants derived requirements for unsafe code from external specifications and the inherent guarantees of the borrow checker. However, they also relied on ad hoc reasoning and tacit knowledge. Many were uncertain if their encapsulations were
sound in all situations. Participants attempted to minimize and
isolate unsafe code, and they avoided using unsafe APIs in favor of safer alternatives. However, this was not always possible in certain domain-specific contexts, such as JIT compilers and operating systems, where unsafe code was used pervasively. Survey respondents who exposed unsafe APIs documented their safety properties, and they relied on documentation to use unsafe APIs correctly. When our participants chose to expose unsafe APIs, they were motivated by performance, ergonomics, and necessity to the same extent as for any arbitrary use of unsafe code. Respondents rarely refactored unsafe code, even if they used it regularly.

\section{Related Work}
\label{section:related}
Here, we discuss relevant results from other studies of unsafe Rust code, comparing our results to quantitative findings where relevant. Most prior studies have examined unsafe code as it is written, either by surveying Rust's package ecosystem~\cite{astrauskas20, evans20, ozdemir16} and standard library~\cite{cui23}, or by examining vulnerabilities caused by unsafe code~\cite{qin20, xu21, zheng23, mirilli, zhang23}. Other studies have collected blogs, articles, and forum posts to understand the implications of unsafe code for the Rust community. Zeng and Crichton~\cite{zeng18} examined a set of articles and comments posted on Hacker News and the Rust subreddit to identify potential challenges to Rust's adoption. Zhu et al.~\cite{zhu22} examined posts on Stack Overflow to determine if developers frequently used unsafe code to resolve compilation errors. These studies provide meaningful contextual information that we can use to cross-validate our findings. However, none can fully answer our research questions.

Far fewer studies have used surveys or interviews to directly engage with developers who use unsafe code. As an extension to their survey of Rust libraries, Evans et al.~\cite{evans20} distributed a short survey on the Rust subreddit. They asked developers why they use unsafe code, which operations they use, and what steps they take to ensure that their code is correct; it received 20 responses. Zeng and Crichton~\cite{zeng18} also interviewed three Rust contributors to provide additional context in their analysis. Fulton et al.~\cite{fulton21} interviewed 16 developers who had advocated for using Rust in industry, followed by a survey that received 178 responses from many of the same communities as ours. Although they used a similar methodology, Fulton et al.~\cite{fulton21} mainly focused on evaluating the benefits and drawbacks of Rust as a general-purpose systems programming language instead of examining Rust's unsafe features. These studies provide evidence that can partially answer a subset of our research questions, vut most do not share our direct focus on unsafe code or do not investigate challenges specific to tool use and interoperation.

Höltervennhoff et al.~\cite{holtervennhoff23} are the only other group that we are aware of that has used qualitative methods to study how developers use unsafe code. Their study was published concurrent to our investigation; after we had completed our interviews but before distributing our survey. They interviewed 26 Rust developers who had experience using unsafe code. Except for three who piloted the interview protocol, all of their participants had contributed unsafe code to open source projects included in the unofficial ``Awesome Rust,'' list~\cite{awesome22} of applications. We cover similar topics, but our research questions reflect different, complementary goals. They focused on understanding how developers reason about the security implications of using unsafe code, while we focused specifically on interoperation and tool use. We provide additional details on challenges associated with tool use and interoperation that are missing from their work, and we demonstrate that their results can generalize to a broader population of developers. 

Here, we describe how our results compare with these prior studies for each of our research questions.

\paragraph{RQ1 - Interoperation} Interoperation has consistently been one of the most common use cases for unsafe code. In 2020, Astrauskas et al.~\cite{astrauskas20} found that 44.6\% of all publicly-exported unsafe functions in the Rust ecosystem were static bindings to foreign functions. Other surveys of the Rust ecosystem have had similar findings; Ozdemir~\cite{evans20} found that 30\% of unsafe blocks were entirely dedicated to foreign function calls, and Evans et al.~\cite{evans20} found that 22.5\% of unsafe function calls were to foreign functions. Zhang et al.~\cite{zhang23} manually analyzed 5,946 unsafe blocks from 140 prominent libraries to determine which unsafe operations were necessary and which could have been avoided. The majority of all raw pointer dereferences and unsafe function calls were marked as necessary to interoperate with foreign libraries. Studies of developers have reached similar conclusions; 70\% of Fulton et al.~\cite{fulton21}'s survey respondents and all but one of Höltervennhoff et al.~\cite{holtervennhoff23}'s interview participants had called foreign functions from Rust. Interoperation was also common among our participants; 70\% of survey respondents had called foreign functions.

However, few studies have identified specific challenges associated with interoperation. Cui et al.~\cite{cui23} explicitly excluded foreign functions from their investigation of Rust's safety properties. Xu et al.~\cite{xu21} found 12 examples of bugs and vulnerabilities that involved foreign function calls. They were related to application-specific invariants, data races, invalid alignment, and inconsistent layouts. However, the authors provided few details on the nature of these bugs, indicating that they were ``straightforward'' and not specific to Rust. Fulton et al.~\cite{fulton21}'s participants had varying experiences with interoperation, but C++ was particularly difficult due to what one participant described as ``rampant aliasing.'' Höltervennhoff et al.~\cite{holtervennhoff23} indicated that some participants struggled with foreign function calls that involved ``memory management,'' but that pure functions were easier to reason about. Our interview participants had similar experiences. Calling pure functions was typically straightforward, as long as the bindings were correct and ABI-compliant. When participants struggled to reason about foreign function calls, it was usually because these APIs had aliasing or concurrency patterns that conflicted with Rust's idioms.

Even fewer studies have identified bugs that meaningfully involved Rust's aliasing model. In 2023, a set of anonymous authors classified 22 bugs from five popular Rust encapsulations of foreign libraries~\cite{anon23}. Most of the errors they investigated involved incorrect handling of exceptions and heap objects that were improperly shared across foreign boundaries. They identified two bugs related to Rust's aliasing restrictions, but they did not describe these bugs in terms of Stacked or Tree Borrows. McCormack et al.~\cite{mirilli} extended Miri to interoperate with an LLVM interpreter and used this hybrid tool to conduct a large-scale evaluation of multilanguage libraries. In Rust, packages and libraries are referred to as ``crates'' and are published on \href{https://crates.io}{crates.io}. They found 46 bugs from 37 crates, and one of these crates was maintained by the Rust Project. This crate had an aliasing violation that occurred after a foreign API had copied a pointer derived from a ``unique'' mutable reference. This fits a pattern described by one of our interview participants, who was concerned that foreign APIs would ``\textit{retain a pointer}'' from Rust and \ilquote{keep operating on it later}{11}. Yu et al.~\cite{yu25} found additional aliasing violations with CapsLock, an instrumentation tool that uses custom RISC-V hardware capabilities. Their approach performs nearly twice as fast as Miri and has support for inline assembly, but its aliasing model is weaker than both Stacked and Tree Borrows, and the capabilities that it relies on have not been implemented in physical hardware yet. They evaluated their design on a variety of test cases that called foreign functions and found similar categories of aliasing violations as McCormack et al~\cite{mirilli}. In particular, certain test cases were ``mutating an owned object without borrowing the owner''~\cite{yu25}.

\paragraph{RQ2 - Tooling} Rust is a relatively new programming language, so most research efforts have focused on developing new tools and not evaluating existing ones. Zeng and Crichton~\cite{zeng18} hypothesized that a lack of promotion for tooling could prevent Rust from being widely adopted as a systems language. However, their work was published prior to the creation of most Rust-specific bug-finding tools. Lack of promotion could explain why few of our participants had used formal methods tools. However, several of the tools listed in our survey, such as Verus~\cite{lattuada23}, had recently been introduced, and few had support for unsafe code. Today, a variety of static verification tools for Rust are being used in industry, and we expect that studying their adoption would be a useful direction for future work. 

Höltervennhoff et al.~\cite{holtervennhoff23} are the only other group that we are aware of to examine how developers use tooling to assist with writing unsafe code. Certain development tools were somewhat underused among their interview population; only seven of their 26 participants had mentioned using Miri. Since 61\% of our survey respondents had used Miri, this does not seem to indicate a broader trend within the Rust community. Likewise, few of their interview participants used automated auditing tools like \code{cargo-audit}, while 51\% of our survey participants had used one or more of these tools. However, our respondents only needed to have used these tools once, while Höltervennhoff et al.~\cite{holtervennhoff23} indicated that participants who audited their dependencies did so ``regularly.'' Höltervennhoff et al.~\cite{holtervennhoff23} also indicated that the majority of their participants had used Clippy, and that some had mixed experiences. Although many of our interview participants had also used Clippy, our investigation was primarily concerned with bug-finding tools, and not linting tools, so we did not highlight it in our analysis or survey. We also cover debuggers, static verifiers, and tools that generate bindings to foreign functions, which were not mentioned in their study. 

\paragraph{RQ3 - Motivations} Multiple studies have identified different use cases for unsafe code. Qin et al.~\cite{qin20} collected random samples of operations within unsafe contexts from Rust's standard library and 10 popular third-party libraries. They identified three use cases for these operations; 42\% were related to interoperation, 22\% were to improve performance, and 14\% involved sharing memory between threads. Astrauskas et al.~\cite{astrauskas20} used ``anecdotal examples'' and snippets of source code to create a taxonomy of six purposes for using unsafe code. These included implementing data structures with complex sharing, overcoming the incompleteness of Rust's type system, emphasizing contracts and invariants, calling foreign functions, using intrinsics or inline assembly, and improving performance. The most frequently used unsafe feature was calling unsafe functions. Fulton et al.~\cite{fulton21} found that developers most frequently used unsafe code to call foreign functions, improve performance, and interact directly with hardware or the operating system. 

Unlike these studies, we are concerned with ``motivations,'' which we define as feature-agnostic reasons for using unsafe code. This excludes several of the feature-specific topics mentioned in prior work, such as intrinsics~\cite{astrauskas20} or sharing memory between threads~\cite{astrauskas20}. We cannot reliably assign motivations to specific features without knowing more about the decision-making process behind each use case. Our perspective is similar to Höltervennhoff et al.~\cite{holtervennhoff23}, who identified necessity, performance, and efficiency as ``motivations.'' It is unclear whether this distinction was also an explicit consideration for their study. Zhang et al.~\cite{zhang23} were able to classify unsafe operations as ``necessary'' by applying a strict definition: it had to be impossible to implement a safe alternative. In their evaluation of Rust libraries, they identified two reasons why developers chose to use unsafe code when it was not strictly necessary: to improve performance, and because they were unaware of a safer alternative. They did not identify whether developers were aware of a safe alternative but had chosen to use unsafe anyway, since it was easier. Their strict criteria also excludes certain context-specific perspectives on what it means for unsafe code to be necessary. For example, they classified all instances of \code{UnsafeCell} as unnecessary, since Rust's safe \code{RefCell} can be used instead. However, it may be necessary to avoid the run-time cost of \code{RefCell} in performance-critical applications. Zhang et al.~\cite{zhang23} account for this by evaluating the performance impact of several safe alternatives to unsafe operations, but we expect that developers' subjective experiences will also be helpful in understanding what motivates them to use unsafe code. 

\paragraph{RQ4 - Encapsulation} Unsafe code is prevalent throughout the Rust ecosystem, but developers use it minimally and hide it under safe interfaces. In July of 2016, Ozdemir~\cite{ozdemir16} found that 29\% of all published crates used unsafe features, but most unsafe blocks were relatively small and tightly-scoped to include only unsafe operations. Evans et al.~\cite{evans20} observed similar patterns in September of 2018. The number of published crates had increased by more than ten times, but 29\% still directly used unsafe code. Although 38\% of crates did not directly use unsafe code, they either directly or indirectly depended on third-party crates that did use these features. In January of 2020, Astrauskas et al~\cite{astrauskas20} found that 23.6\% of third-party crates used unsafe features. Most instances of unsafe code were relatively straightforward and hidden beneath safe interfaces. The majority of unsafe functions called by libraries were standard (e.g. not closures, function pointers, or trait functions) and defined within the same crate or provided by Rust's standard library. In particular, 34.7\% of all crates had declared unsafe APIs, but they were not visible to users. This aligns with our participants' experiences; 88\% claimed that they would attempt to encapsulate unsafe operations beneath safe APIs, and 61\% expected that it would be at least somewhat easy for another developer to understand a random unsafe block or function from their code. 

Developers typically mark functions with the \code{unsafe} keyword to indicate that users need to uphold a particular safety property. The majority (61\%) of the survey respondents who had exposed unsafe APIs only did so when they required an invariant beyond what Rust's type system could provide. However, Astrauskas et al.~\cite{astrauskas20} found that 36.1\% of unsafe functions had completely safe implementations. On further investigation, many of these functions had been automatically generated or annotated as \code{unsafe} for ``legacy reasons.'' This could provide an alternative explanation beyond necessity or ease-of-use for why the remaining 39\% of our respondents were motivated to expose unsafe APIs. While unsafe functions are relatively common, developers rarely use unsafe traits. Evans et al.~\cite{evans20} found that only slightly more than 1\% of all crates declared an unsafe trait, and only 6\% implemented an unsafe trait. Astrauskas et al.~\cite{astrauskas20} had similar findings; only 2.5\% of all traits were unsafe, and 40.4\% of these declarations originated from five crates. Likewise, Zhang et al.~\cite{zhang23} only identified 11 instances of unsafe trait implementations. Astrauskas et al.~\cite{astrauskas20} suggested that this lack of use could be caused by a lack of guidance on when a trait should be labeled \code{unsafe}. One of our interview participants indicated that certain developers may have misconceptions on the role of unsafe traits. As mentioned earlier in~\tlink{Section}{results:rq4}, one participant who maintained a library that exposed an unsafe trait found that users who had implemented it incorrectly would blame the library, even though \ilquote{in theory, the burden of proof is on them}{7}. Both the Rust Reference~\cite{rustref} and the Rustonomicon~\cite{nomicon} have brief sections on unsafe traits, but additional resources could be helpful.

The safety properties for unsafe features are varied and complex. Cui et al.~\cite{cui23} manually reviewed 416 unsafe APIs in Rust's standard library and identified 19 categories of preconditions and postconditions. Partly for this reason, discrepancies between the signatures of safe APIs and their interior unsafe implementations have been a significant source of bugs and vulnerabilities for the Rust ecosystem. Qin et al.~\cite{qin20} examined 170 bugs and vulnerabilities from ten high-profile Rust libraries. Memory safety errors typically occurred in safe contexts but were caused elsewhere by errors in interior unsafe implementations. Qin et al.~\cite{qin20} also identified 19 examples of incorrect encapsulations of unsafe code from both the third-party libraries in their sample and within Rust's standard library. Xu et al.~\cite{xu21} exclusively studied registered vulnerabilities and identified similar errors in safe encapsulations. Zheng et al.~\cite{zheng23} conducted the broadest empirical study of security issues in the Rust ecosystem to date. They examined 433 unique vulnerabilities that were discovered between 2014 and 2022, identifying 300 affected repositories. Vulnerable components were significantly more likely to include unsafe blocks or functions. The overwhelming majority of all vulnerabilities affected safe functions.

For this reason, it is somewhat concerning that our participants were often uncertain about the correctness of their encapsulations. Höltervennhoff et al.~\cite{holtervennhoff23} had similar findings; most of their participants relied on common sense and test coverage to evaluate the security of their code, and few had contributed to projects with formal policies for security or code review. However, many of their participants indicated that they would review unsafe code more carefully to ensure that it upheld all necessary safety invariants. Few participants had ever encountered security issues caused by unsafe code, and none of these issues were critical. Many of our interview participants had encountered memory or thread-safety bugs in unsafe implementations, but few described the security implications of these issues. Security was not a central focus of our study, so it is possible that our participants had reflected on the security implications of their decisions but neglected to mention this during interviews.

\section{Discussion}
\label{section:discussion}
Our goal in studying interoperation was to describe how developers navigate the differences between Rust and other languages (\hyperref[rq1]{RQ1}) and to determine how tooling could be improved to make this process easier (\hyperref[rq2]{RQ2}). Our interview participants found that interoperation was more or less difficult depending on how the characteristics of foreign APIs matched the inherent expectations of Rust's aliasing model. When documentation was missing or certain design idioms conflicted with Rust's aliasing rules, participants had no choice but to minimize their interaction with foreign function calls to reduce the perceived risk of triggering undefined behavior. At the time this survey was conducted, Miri was the only tool capable of finding violations of Rust's aliasing model---it was both the most popular tool and the most lamented. Participants indicated that they had been deterred from using it due to its lack of support for foreign function calls and inline assembly, or its slow performance.

These challenges are not localized at foreign function boundaries. Developers need to reconcile the semantics of foreign APIs with safe, idiomatic Rust encapsulations, which often requires implementing a layer of interior-unsafe operations. Developers' motivations for using these operations (\hyperref[rq3]{RQ3}) and their ability to reason about encapsulating unsafe code (\hyperref[rq4]{RQ4}) are inherently relevant to this process. Our population prioritized exposing safe encapsulations whenever possible. Necessity was the most predominant motivation for exposing unsafe APIs, and the overwhelming majority (\ref{encap:unsafe-api:docs}) of participants indicated that they would always document any necessary preconditions and postconditions. However, factors that are orthogonal to correctness, like ergonomics and performance, influenced how participants used unsafe code within safe encapsulations. For example, one interview participant indicated that it had been easier to encapsulate a foreign API by transmuting a pointer to a long-lived allocation into a reference with a \code{{\textquotesingle}static} lifetime. The alternative would have been to derive a context-specific lifetime, and another participant indicated that this pattern could make an API more difficult to use. These motivations were less common but still significant---51\% of participants who exposed unsafe APIs cited performance as a motivation.

We expect that Rust developers will continue to prioritize exposing safe APIs, at least in part due to community incentives. More than 80\% (\ref{encap:avoid}) of participants who had used an unsafe APIs before would prefer a safe alternative at least most of the time, if not always. As one interview participant put it: \ilquote{everyone wants a safe interface... no one really wants to go use the unsafe heavy ones}{9}. This may explain why interview participants were willing to rely on ad hoc, application-specific reasoning to justify their design decisions in the face of domain-specific challenges and limitations in tooling. As a result, developers are left uncertain about their design decisions. Only 14\% (\ref{motivation:necessity}) of survey respondents who used unsafe code by necessity were always certain that it would be impossible to implement a safe alternative. Likewise, 62\% (\ref{q:uncertain:safe-encap}) of respondents who exposed safe APIs for unsafe code were certain ``most of the time'' that their interfaces were sound, but only 23\% were always certain.

Participants suggested several potential solutions to this uncertainty. Developers who contributed to applications with a significant body of specialized unsafe features---mainly multilanguage applications, JIT compilers, and operating systems---would benefit from new dynamic analysis tools for finding Rust-specific aliasing bugs at-scale. Static analysis could also be helpful to generate and validate bindings and encapsulations for unsafe functions. However, Rust is also constantly evolving, and interview participants identified several topics that they were uncertain about due to Rust not having a formal, comprehensive specification. For this reason, we expect that our population would benefit from more comprehensive documentation on certain unstable language features. 

\subsection{Dynamic Tooling}
Our participants also indicated that they needed new tools for finding Rust-specific forms of undefined behavior: \ilquote{a version of Miri that could go across that FFI boundary}{18}, or \ilquote{...a unique pointer sanitizer}{1}. Miri has experimental support for executing foreign functions from shared libraries. However, it is still several orders of magnitude slower than native execution, prohibiting it from being useful in the types of large-scale C and C++ applications where Rust is increasingly being adopted. We see two potential routes for implementing new tools in this category. The first would be to insert run-time checks for aliasing violations during compilation. This would require lowering a subset of Rust's type information from the MIR level down into the intermediate representations used by these platforms, so that the instrumentation pass could distinguish between references and raw pointers. Yu et al.~\cite{yu25} have had success with this method in CapsLock, but their approach uses an aliasing model that is strictly weaker than Stacked and Tree Borrows. In ongoing work, we are creating BorrowSanitizer~\cite{bsan}, which is implemented without hardware acceleration, allowing us to have parity with Miri's aliasing models. Our approach uses many of the same APIs as other popular LLVM-based tools.

We expect that compile-time instrumentation tools will provide the combination of performance and compatibility that is necessary to support large-scale, multilanguage applications. However, participants who contributed to JIT compilers would also benefit from dynamic binary instrumentation. Tools in this category, such as Valgrind~\cite{seward07}, can encode and instrument native assembly on-the-fly, which would be necessary to support JIT-compiled assembly. In 2023, the Krabcake project~\cite{krabcake} proposed extending Valgrind to be capable of finding Stacked Borrows violations. Several participants had mentioned Krabcake and indicated that it would be a natural extension to their workflows, since they had already used Valgrind with Rust programs. It was also the second most popular dynamic bug-finding tool behind Miri; it was used by 44\% of survey respondents. However, Krabcake is no longer actively maintained, and it is not yet capable of finding bugs in real-world programs. The Rust community should consider reactivating this project.

\subsection{Static Tooling}
Participants also indicated that static analysis tools could be useful for finding undefined behavior across foreign function boundaries. Several analyses could be implemented as extensions to binding generation tools at varying levels of sophistication. One participant wanted a tool that could tell \ilquote{whether your FFI conforms to your expectations...Do all my symbols match? Do all the types signatures match?}{5}. McCormack et al.~\cite{mirilli} were able to detect errors in foreign function bindings by comparing the size of types in the Rust declarations with the size of types in the LLVM definitions. The authors implemented this as part of a dynamic analysis tool, but their method did not depend on dynamic information, so it could be factored out as a static analysis pass. Changes in Rust's newest 2024 edition indicate that Rust will continue to evolve toward improving contract visibility at foreign APIs. Rust's \code{extern} blocks now require \code{unsafe} annotations~\cite{unsafe_extern}, indicating that developers are responsible for ensuring that foreign function bindings match their definitions. Another participant tried to avoid manually reasoning about the lifetime semantics of foreign APIs \ilquote{in hopes that in future, there will be a better method for doing this correctly}{4}. In 2022, contributors to Clang proposed adding lifetime annotations to C++, which would be inferred through a whole-program static analysis pass~\cite{lifetime_inference}. The particular RFC  has not had activity since August 2023, but the project was associated with Google’s Crubit analysis tool, which is still in development as of May 2025~\cite{crubit}. 

Despite their uncertainty, few of our participants had tried using static verification tools. One possible explanation is that when we distributed the survey, only Verus~\cite{lattuada23} and Kani~\cite{kani} had significant support for verifying unsafe code. Since then, both of these tools have added extensive support for unsafe features. Verus has been used to verify Vest~\cite{vest}, a parser combinator library, as well as several other security-critical application components~\cite{verus_project_page}. Both Kani and Verifast~\cite{jacobs24} are being used to verify the entirety of Rust's standard library as part of a joint initiative between Amazon and the Rust Foundation~\cite{verify_std_announcement, verify_std}. We expect that these new tools will meet our participants' needs for stronger guarantees of soundness. For example, two of our interview participants indicated that refinement typing could be useful---\ilquote{custom subtyping, so that you can specify I want an integer, but only between one and ten}{4}. Flux~\cite{lehmann23}, a refinement-type checker for Rust, was published concurrent to our study and has since been used to verify several real-world Rust applications~\cite{lehmann25, vtock}. As verification continues to become standardized within the Rust community, it would be worthwhile for future studies to examine how these methods are being deployed in practice.

\subsection{Unsafe Code Guidelines} 
The majority of survey respondents found that the Rust community's guidance and documentation were helpful, but only 6\% (\ref{q:uncertain:docs}) indicated that these resources were \textit{always} enough to determine how to use unsafe code correctly. Interview participants also expressed uncertainty about Rust's semantics. Some were uncertain about the interaction between Rust's \code{panic!} macro and \code{Drop} implementations, with one participant indicating that \textit{``real guidelines for library authors''} would be helpful in this area. This has been a subject of active discussion since at least 2021~\cite{panic_drop_nomicon}, but has yet to be fully documented in community resources~\cite{panic_docs, panic_drop_nomicon}. We expect that other, similar topics lack formal documentation—instead, developers need to examine years-long issue threads within repositories for Rust's compiler and Unsafe Code Guidelines~\cite{ucgref}.

We agree with Höltervennhoff et al.~\cite{holtervennhoff23} that developers would benefit from having a dedicated resource summarizing points of contention about Rust's semantics. However, Rust is still a relatively new systems programming language, and certain feature interactions have yet to be fully explored or agreed-upon to the point where writing adequate documentation would even be possible. Developers who implement design patterns that sit at the cutting-edge of Rust's evolving semantics will inevitably be uncertain if their implementations are sound. Even so, it would be worthwhile for the Rust team to continue to review popular, long-standing issue threads and flag topics that have not been fully documented in the Rustonomicon~\cite{nomicon} or the Rust Reference~\cite{rustref}. Meanwhile, developers who engage with unsafe code should periodically review relevant issues within the Unsafe Code Guidelines repository to ensure that they are using these features correctly and to have the opportunity to inform compiler developers about their use-cases. However, we still expect that improvements to Miri will have the greatest impact on our survey population, since it was used by a majority of survey respondents, and it serves as somewhat of an executable specification for Rust. 

\section{Conclusion}
\label{section:conclusion}
The Rust programming language provides static safety guarantees. However, Rust is frequently used to interoperate with other languages. To call foreign functions, developers need to use Rust's unsafe features, which bypass its safety restrictions. These features can introduce bugs and security vulnerabilities if they are used incorrectly. Few Rust-specific development tools provide comprehensive support for unsafe code, and even fewer support multilanguage applications. Developers will need new solutions for testing, bug finding, and verification in this context. We conducted a mixed methods, exploratory study to determine which interventions will be effective for developers who use unsafe code. We interviewed 19 developers who regularly wrote or edited unsafe code and created a survey that reached an additional 160 developers who engaged with unsafe code in some capacity. We examined the challenges associated with interoperation, limitations in tooling, developers' motivations for using unsafe code, and how they reason about encapsulation. 

The majority of participants used unsafe code to call foreign functions. Certain foreign libraries had aliasing and concurrency patterns that conflicted with the expectations of Rust's type system, making them difficult to encapsulate. However, participants did not have a reliable method to determine if their encapsulations were sound. Miri, a Rust interpreter, was the most popular solution for detecting undefined behavior, but participants who had used it before were deterred from using it again because of its slow performance and lack of support for key features. Most participants were motivated to use unsafe code by necessity, but they also cited performance and ergonomics as motivations. The majority followed best practices, such as encapsulating and documenting unsafe code, but many were somewhat uncertain about their design patterns, and few felt that Rust's documentation was always adequate for understanding how to use unsafe code correctly.

Our participants would benefit from new static and dynamic analysis tools to assist with interoperation and with finding Rust-specific forms of undefined behavior. Dynamic analysis tools could improve on Miri to provide better performance, support for inline assembly, and the capability to detect aliasing violations triggered by operations that span foreign function boundaries. The Rust community should also continue to invest in improving and expanding documentation on unsafe code, so that longstanding issue threads can become integrated into standard community resources. A useful direction for future work would be to synthesize the methods used by Astrauskas et al.~\cite{astrauskas20} and Evans et al.~\cite{evans20}, which provide different, complementary perspectives on the distribution of unsafe code. Future studies should consider the use of unsafe code in unpublished but publicly available crates, as well as the extent to which Rust libraries are exposed to unsafe operations through foreign function calls. The results of such a study would provide an excellent opportunity to triangulate the results described in this paper.

\begin{acks}
We thank the Rust Community for their engagement and support, as well as Jenny Liang and Courtney Miller for their feedback on our study design.
\end{acks}

\bibliographystyle{ACM-Reference-Format}
\bibliography{refs}
\end{document}